\documentclass[aps,pra,twocolumn,superscriptaddress,amsmath,amssymb]{revtex4-2}

\usepackage[utf8]{inputenc}
\usepackage{amsmath,amsthm}
\usepackage{graphicx}
\usepackage[colorlinks = true,
            linkcolor = blue,
            urlcolor  = blue,
            citecolor = blue,
            anchorcolor = blue]{hyperref}
\usepackage{braket}
\usepackage{xcolor}
\usepackage{easyReview}
\usepackage{bm}
\newcommand{\tr}[1]{\ensuremath{\operatorname{tr}\!{#1}}}

\newcommand{\imag}{{\rm Im}}
\newcommand{\bea}{\begin{eqnarray}}
\newcommand{\eea}{\end{eqnarray}}

\newcommand{\BB}{\mathbb{B}}

\def\RE{\mathop{\rm Re}} %
\def\IM{\mathop{\rm Im}} %

\global\long\def\ga{\gamma} \global\long\def\de{\delta}
\global\long\def\De{\Delta} \global\long\def\Ga{\Gamma}
\global\long\def\th{\theta}
\global\long\def\th{\theta}

\global\long\def\ell#1{\theta_{#1}}
\global\long\def\bell#1{\tilde\theta_{#1}}

\global\long\def\la{\lambda} \global\long\def\ka{\kappa}
\global\long\def\si{\sigma}
\global\long\def\vfi{\varphi}

\global\long\def\eps{\epsilon}
\global\long\def\al{\alpha}
\global\long\def\be{\beta}
\global\long\def\ga{\gamma} \global\long\def\de{\delta}

\def\brho{\bm{\rho}}

\global\long\def\no{\nonumber}

\newtheorem{hyp}{Hypothesis}

\theoremstyle{remark}

\global\long\def\braket#1#2{\left\langle #1|#2\right\rangle }

\renewcommand{\thefigure}{\arabic{figure}}

\global\long\def\bell#1{\theta_{#1}}
\global\long\def\ell#1{\bar{\theta}_{#1}}

\begin{document}

\title{Dissipatively dressed  quasiparticles  
in boundary driven integrable spin chains }

\author{Vladislav Popkov} \affiliation{Faculty of Mathematics and
  Physics, University of Ljubljana, Jadranska 19, SI-1000 Ljubljana,
  Slovenia} \affiliation{Department of Physics, University of
  Wuppertal, Gaussstra\ss e 20, 42119 Wuppertal, Germany}
\affiliation{ These two authors contributed equally to this work}

\author{Xin Zhang} \affiliation{Beijing National Laboratory for
  Condensed Matter Physics, Institute of Physics, Chinese Academy of
  Sciences, Beijing 100190, China} \affiliation{ These two authors
  contributed equally to this work}

\author{Carlo Presilla} \affiliation{Dipartimento di Matematica,
  Universit\`a di Roma La Sapienza, Piazzale Aldo Moro 5, Roma 00185,
  Italy} \affiliation{Istituto Nazionale di Fisica Nucleare, Sezione
  di Roma 1, Roma 00185, Italy}

\author{Toma\v z Prosen} \affiliation{Faculty of Mathematics and
  Physics, University of Ljubljana, Jadranska 19, SI-1000 Ljubljana,
  Slovenia} \affiliation{Institute of Mathematics, Physics and
  Mechanics, Jadranska 19, SI-1000 Ljubljana, Slovenia}

\begin{abstract}
The nonequilibrium steady state (NESS) of integrable spin chains experiencing strong boundary dissipation
is accounted by introducing  quasiparticles with a renormalized—dissipatively
dressed—dispersion relation. This allows us to evaluate the spectrum of the NESS in terms of the Bethe ansatz equations for a related coherent system which has the same set of eigenstates, the so-called dissipation-projected Hamiltonian.
We find explicit analytic expressions for the dressed energies of the XXX and
XXZ models with effective, i.e., induced by the dissipation, diagonal boundary fields, which are
U(1) invariant, as well as the XXZ and XYZ models with effective non-diagonal boundary fields. In 
all cases, the dissipative dressing generates an extra singularity in the dispersion relation, 
substantially altering the NESS spectrum
with respect to the spectrum of the corresponding coherent model.
\end{abstract}

\maketitle

\section{Introduction}

The notion of quasiparticles is a fundamental concept in regular many-body systems irrespectively on whether they are integrable or not.  Examples of quasiparticles are phonons in crystals and magnons in magnetic materials.  In integrable continuous theories 
with nonlinearities
(e.g. in Korteweg de Vries or sine-Gordon models),  stable quasiparticles are solitons, or solitary waves which are spatially localized:   
after a collision they  retain their original velocities and shapes,   the net result of the scattering being just a shifting 
of their world lines.   
Quasiparticles of integrable quantum interacting manybody systems on a  lattice  
 also possess special properties with regard of their multiple scattering:
the multiple quasiparticle collision can be factorized in terms of two-body  scattering matrix.  
Every eigenstate in an integrable quantum system can be viewed as consisting of  quasiparticles.  Most prominently, 
the eigenenergy of any eigenstate  is 
 obtainable by summing individual  energies of all quasiparticles it contains $E_\al = \sum_j \eps (u_{j,\al})$
where $u_{j\al}$ are set of admissible rapidities and $\eps (u)$ is a dispersion relation. The 
dispersion relation $\eps (u)$  is a fundamental  property of  quasiparticles,  and it is very robust,  e.g.  
the dispersion does not depend on boundary conditions,  temperature,  system size  etc.
 
The purpose of our communication is to establish a relation between two fundamental quantities from seemingly very 
different fields: the spectrum $E_\al$  of a coherent (closed) integrable many-body system,  from one side,  and 
the spectrum $\la_\al$  of a non-equilibrium steady state of the ``same'' open quantum system in contact with locally acting dissipative 
bath, from another side.   In more details,  we will demonstrate that under a proper choice of the dissipation,  there exists
a one-to one correspondence between the two spectra,  
namely that  $\log \la_\al =  \sum_j \tilde{\eps} (u_{j\al})$ where $u_{j\al}$ is exactly the same sets of admissible rapidities,  constituting the spectrum of a coherent system $E_\al = \sum_j \eps (u_{j\al})$,  and    $\tilde{\eps}(u)$ is a renormalized 
dispersion (we refer to it as {\em dissipatively dressed dispersion relation}).  An existence of a renormalized dispersion relation 
relating coherent (unitary) and strongly incoherent (dissipative) systems seems rather counterintuitive; however its  
general validity lies beyond what we are currently able to prove.


We compare an equilibrium Gibbs state $\rho_\mathrm{Gibbs}$  of a quantum integrable many-body system, from one side,  and a nonequilibrium steady state 
$\rho_\mathrm{NESS}$
of ``almost the same"  system,  with a part of degrees of freedom 
coupled to  a dissipative bath,  from the other side. 
The Gibbs state is
\begin{align}
  \begin{split}
    &\rho_\mathrm{Gibbs} = \frac{1}{Z}
      \sum_{\al}  e^{-\be E_\al} \ket{\al}  \bra{\al},
    \\
    &E_\al = \sum_j \eps (u_{j,\al}),
  \end{split}
  \label{eq:rhoGibbs}
\end{align} 
where $\eps(u)$ is the dispersion relation of the quasiparticles, and
$u_{j\al}$ are the admissible rapidities, obtained via the
quantization conditions, namely, a set of Bethe ansatz equations
(BAE) obtainable by different methods ~\cite{1931Bethe,BaxterBook,1979Takhtajan,1980QISM,1995SkyaninSOV}.
The type of quasiparticles and their number depend on the
model and the intrinsic quantum numbers characterizing the microstates
$\{\ket{\al}\}$.

In the dissipative setup,  we  consider a ``similar" quantum system, with boundary degrees of freedom 
coupled to  a dissipative bath, described via a Lindblad Master equation.  
This  leads to a nonunitary dynamics of a reduced density matrix $\rho(t)$,  which  relaxes to a (unique)  nonequilibrium steady state (NESS),
$\lim_{t\rightarrow \infty}\rho(t) = \rho_{{\rm NESS}}$. 
In general,  there  is no reason to expect any close relation between the Gibbs state (\ref{eq:rhoGibbs}) and the nonequilibrium 
NESS. 

The purpose of this communication is to show,  that under certain conditions, (A)  the NESS has  the same 
set of eigenstates $\ket {\al} $ as (\ref{eq:rhoGibbs}),  i.e. 
\begin{align}
[ \rho_{{\rm NESS}},\rho_\mathrm{Gibbs}] =0, \label{eq:NESSgibbsComm}
\end{align}
  and, (B) the effective energies  $\tilde{E}_\al$ are described 
by exactly the same quasiparticle content as  $E_\al$, i.e.  are given by
the sum of the same set of quasiparticles as in (\ref{eq:rhoGibbs}),  but with renormalized---``dissipatively
  dressed"---dispersion relation $\eps(u) \rightarrow \tilde \eps(u) $, 
\begin{align}
  &\rho_{{\rm NESS}}  =
      \frac{1}{\ \tilde{Z}}
      \sum_{\al} e^{- \tilde{E}_\al}  \ket{\al}  \bra{\al},\label{eq:NESS}\\
 &  \tilde{E}_\al =  \sum_j \tilde {\eps}(u_{j,\al}). \label{eq:dissDressing}
\end{align} 
The relation (\ref{eq:dissDressing}) is very surprising:  it  suggests that the notion of Bethe quasiparticles (a 
intrinsic property of a coherent integrable system) can,  under certain conditions,  be preserved
under dissipation,  at least with respect to its steady state (NESS). 

 The following sections serve to explain the relation between coherent and dissipation-affected quantum systems in detail. 
 
 We note that this is a companion paper expanding on a recent exact result~\cite{2024NESSspecArxiv}, here including heuristic and numerical evidence in a broader family of integrable models and dissipation setups.

\section{Integrable spin chains with boundary fields and
boundary-driven spin chains}
\label{sec:EffectiveDynamics}

Our coherent system is a general open spin $\frac12$ chain described by nearest neighbour Heisenberg Hamiltonian with general boundary fields $h_1=\vec{h}_1\cdot\vec{\si}_1$ and $h_N=\vec{h}_N\cdot\vec{\si}_N$, 

\begin{align}
  &H = H_{\rm bulk}  +h_1 + h_N \label{eq:OpenXYZ}\\
  &H_{\rm bulk}= \sum_{n=1}^{N-1} \vec{\si}_{n} \hat {J} \vec{\si}_{n+1}, \quad \hat{J} = \mathrm{diag}(J_x,J_y,J_z).
  \label{eq:Hbulk}
\end{align}
It is  known,  see \cite{2014Cao,CaoBook}, that   the above system is  integrable for any choice of $J_x,J_y,J_z$, $\vec{h}_1,\vec{h}_N$.  
More popular examples of XXZ and XXX model result from reductions  $J_x=J_y$ and 
$J_x=J_y=J_z$.  As such,  any eigenstate of (\ref{eq:OpenXYZ})  has a quasiparticle 
content,  with admissible values of the rapidities in (\ref{eq:rhoGibbs}),  given by 
the respective Bethe Ansatz equations \cite{CaoBook}.  Later on, the boundary fields will be specified.

\begin{figure}
  \centering
 \includegraphics[width=0.95\columnwidth,clip]{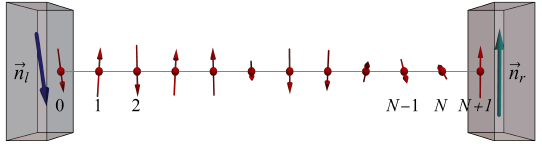}
  \caption{ Schematic picture of the dissipative setup.  The boundary spins are fixed by dissipation, while the 
internal spins follow effective dynamics (\ref{LMEeff}),  consisting of
 fast coherent  dynamics  (\ref{eq:OpenXYZ}) and slow relaxation dynamics (\ref{app:ClassicalNESS}) towards the NESS.}
  \label{FigSchema}
\end{figure}

On the dissipative side,  our  scenario is as follows: 
we have a spin chain of $N+2$ sites (note that there are two extra sites w.r.t.  the coherent system above),  described by the Hamiltonian 
 $H =\sum_{n=0}^{N} \vec{\si}_{n} \hat {J} \vec{\si}_{n+1}$ (the same Hamiltonian as in (\ref{eq:Hbulk}) with $2$ sites added),  in which the edge spins,  those at positions $n=0$ and  $n=N+1$ are projected on pure qubit states $\rho_l, \rho_r$ respectively,  by coupling to a dissipative Lindblad bath.  Schematical picture is given in Fig.\ref{FigSchema}.

The time evolution of the reduced density matrix is described by the Lindblad Master equation

\begin{align}
&\frac{\partial \brho(\Ga,t)}{\partial t} = - i [H,\brho] + \Ga \left( {\cal D }_{L_1}[\brho] +  {\cal D }_{L_2}[\brho]  \right)\label{LME}\\ 
&{\cal D }_{L}[\brho] = L \brho L^\dagger - \frac12 \left(  L^\dagger L \brho + \brho  L^\dagger L \right). \label{LindbladOperator}
\end{align}

The Lindblad operators can be conveniently chosen in a form 
$L_1 = \ket{\psi_l}\bra{\psi_l}^\perp \otimes I^{\otimes_{N+1}}$,
$L_2=  I^{\otimes_{N+1}} \otimes \ket{\psi_r}\bra{\psi_r}^\perp $,  where 
$\braket{\psi^\perp}{\psi}=0$.  
Indeed the  Lindblad dissipators ${\cal D }_{L_1}$  and  ${\cal D }_{L_2}$  
will then promote the relaxation of edge spins to 
  targeted pure states $\ket{\psi_{l,r}}$, with typical relaxation time $\tau_{\rm boundary} =O(1/\Ga)$.
 We will  consider the case when $\tau_{\rm boundary}$ is much smaller than a typical time needed for the bulk relaxation $\tau_{\rm boundary}\ll \tau_{\rm bulk}$, the so-called quantum Zeno (QZ) regime. Note that $\Ga$ need not be necessarily very large to reach QZ; e.g. for XXX bulk Hamiltonian $J_x=J_y=J_z$
the effective $\Ga=O(N^{-1})$ as shown in \cite{MPA-PRA2013}. 
 
The quantum Zeno regime with arbitrary $\Ga$ can be realized by a protocol of repeated interactions.    
In this protocol,  the edge spin is repeatedly brought in contact with a freshly  polarized magnet, see e.g.\cite{2014Karevski}. 
Alternatively,  an instantaneous  projection of a single qubit can be achieved
 by  applying of a so-called reset  gate  ~\cite{2024Abanin}: the gate simply resets a qubit state to a desired 
qubit state  at each application. In Appendix~\ref{app:ZenoBrickwall} we give a derivation of
 (\ref{LME}) from a qubit circuit protocol. 

The  fast relaxation of the edge spins in QZ regime constrains the reduced density matrix of (\ref{LME}) to an approximately factorized form \cite{2014Zanardi} $\brho(\Ga,t) = \rho_l \otimes \rho(\Ga,t) \otimes \rho_r + O(1/\Ga)$ for $t\gg 1/\Ga$,  where $\rho(t) \approx {\rm tr}_{0,N+1} \brho(t)$  is the reduced density matrix after the tracing over the 
part directly affected by the dissipation \cite{2014Zanardi},  i.e. the reduced density matrix of 
interior spins.

Moreover,  it can be shown that $\rho(\Ga,t)$ \textit{commutes} with the Hamiltonian 
(\ref{eq:OpenXYZ}) for suitably chosen boundary fields $h_1,h_N$  \cite{2018ZenoDynamics},  confirming that the time
evolution of $\brho(\Ga,t) $  for $t\gg 1/\Ga$ is approximately coherent \cite{2014Zanardi}. 

For our purposes it is enough to demonstrate that $\lim_{\Ga \rightarrow \infty} \lim_{t \rightarrow \infty } \rho(t)= \rho_{\rm NESS}$ satisfies 
\begin{align}
&[\rho_{\rm NESS}, H_{\rm bulk} +h_1+h_N]=0 \label{eq:Comm}\\  
&h_1 = \vec{\si}_1 \cdot \hat J \vec n_l \equiv \sum_{\al = x,y, z} J_\al n_l^\al \si_1^\al \label{eq:h1}\\
&h_N = \vec{\si}_N \cdot \hat J \vec n_r , \label{eq:hN}
\end{align}
($H_{\rm bulk}$ given by (\ref{eq:Hbulk})),
so we shall sketch a proof here.  To that end,  we expand the steady state solution $\brho(\Ga)$ of (\ref{LME}) in powers of $1/\Ga$,
$\brho(\Ga) =\rho_0 + \Ga^{-1} \rho_1 + \Ga^{-2} \rho_2+\ldots$  Substituting into (\ref{LME}) we obtain a set of 
conditions $i [H, \rho_k]=({\cal D }_{L_1}+  {\cal D }_{L_2})[\rho_{k+1}]$,  for $k=0,1,\ldots$.  These conditions are nontrivial since the operator $({\cal D }_{L_1}+  {\cal D }_{L_2})$ has nonzero kernel of the form $\rho_l \otimes X \otimes \rho_r$ where 
$X$ is an arbitrary matrix.  In all orders, the necessary and sufficient condition for the existence of $\rho_{n+1}$ is 
${\rm tr}_{0,N+1} [H,\rho_{n}] =0$.  In the leading order $n=0$,  a substitution of $\rho_0 = \rho_l \otimes \rho_{\rm NESS} \otimes \rho_r$   into ${\rm tr}_{0,N+1} [H,\rho_{0}] =0$ leads to    (\ref{eq:Comm})-(\ref{eq:hN}).  The Hamiltonian (\ref{eq:Comm}) is called a dissipation-projected 
Hamiltonian \cite{2014Zanardi}.  Thus we obtain the property (A) (\ref{eq:NESSgibbsComm}).

\textit{Remark 1.~} We proved that the eigenstates of Zeno NESS for boundary-driven integrable spin chains are given
by integrable Hamiltonian with fine-tuned boundary fields   $h_1, h_N$.  Note 
that  due to (\ref{eq:h1}), (\ref{eq:hN}) the conditions 
\begin{align}
&\min_\al J_\al \leq \| h_1\|,\quad \|h_N\| \leq\max_\al J_\al
 \end{align}
are always satisfied,  restricting  the norm of ``permissible" boundary fields in (\ref{eq:Comm}).

To show the property (B)  is more involved as it requires,  to start with,   a description of the relaxation mechanism of internal spins 
towards the steady state, 
 $\rho(t) \rightarrow
\rho_{\rm NESS}$ for $t\gg 1/\Ga$.  This needs
writing down a Dyson expansion of the time-dependent solution of the Lindblad Master equation (\ref{LME}) using 
$1/\Ga \ll 1$ as a perturbation parameter.  The procedure is explained  in detail in  \cite{2014Zanardi,2018ZenoDynamics},
while here we give an  outline.  
The first two orders of the Dyson expansion yield an  effective
Lindblad Master equation for internal spins $\rho(t) = \mathrm{tr}_{0,N+1} \brho(t)$ for $t\gg 1/\Ga$
\cite{2018ZenoDynamics},
\begin{align}
&\frac{\partial \rho(\Ga,t)}{\partial t} = - i [H_{D},\rho] + \frac{1}{\Ga} \left( {\cal D }_{g_l}[\rho] +  {\cal D }_{g_r}[\rho]  \right),\label{LMEeff}
\end{align}
where $H_D \equiv H_{\rm bulk} +h_1+h_N$ is the  dissipation-projected Hamiltonian, and $g_l,g_r$ are effective Lindblad operators, 
explicitly given by:
\begin{align}
&g_l= \mathrm{tr}_{0,N+1} \left[ \left(\ket{\psi_l} \bra{\psi_l^\perp} \otimes I^{\otimes_{N+1}} \right) H \right]\no \\
&=  \mathrm{tr}_{0} \left[ \bra{\psi_l^\perp}  \vec{\si} \ket{\psi_l}  \otimes \hat {J} \vec{\si}\right] \otimes I^{\otimes_{N-1}},\label{eq:gL}\\
&g_r= \mathrm{tr}_{0,N+1} \left[ \left(  I^{\otimes_{N+1}} \otimes \ket{\psi_r} \bra{\psi_r^\perp}\right) H \right] \no\\
&=   I^{\otimes_{N-1}} \otimes \mathrm{tr}_{1} \left[ \hat {J} \vec{\si} \otimes  \bra{\psi_r^\perp}  \vec{\si} \ket{\psi_r}   \right]. \label{eq:gR}
\end{align}
Note that  the dissipative part in  (\ref{LMEeff}) is  of the order $1/\Ga$,  in contrast to (\ref{LME}) where it is of order $\Ga$.
Thus, indeed,  the effective time evolution of the bulk is approximately coherent while the relaxation is proportional to $1/\Ga$.

Due to (\ref{eq:Comm}) the NESS  can be written in the basis of eigenstates  $\ket{\al}$ of $H_{\rm bulk}+ h_1+h_N$,
\begin{align}
&\rho_{\rm NESS}  = \sum_\al \nu_\al \ket{\al} \bra {\al}.  \label{app:NESS}
\end{align}
Assuming orthonormality  $\braket{\al}{\be}= \de_{\al \be}$,  and multiplying 
(\ref{LMEeff}) by $\bra{\al}$ from the left and by $\ket{\al}$ from the right,
one obtains a closed system for the populations  $\nu_\al(t)= \bra{\al} \rho(t) \ket{\al}$:
  \begin{align}
&\Ga  \frac{\partial \nu_\al(t)}{\partial t} = \sum_{\be \neq \al} w_{\be \al} \nu_\be
  - \nu_\al \sum_{\be \neq \al} w_{\al \be},
  \quad \al=1,2,\dots 
  \label{app:ClassicalNESS}\\
 &  w_{\be \al} =
  |\bra{\al} g_l \ket{\be}|^2 +  |\bra{\al} g_r \ket{\be}|^2.
  \label{app:rates}
\end{align}
Quantities $\nu_\al(t) = \bra{\al} \rho(t) \ket{\al}$ are real,  nonnegative and sum up to $1$,
(due to normalization $\tr \rho(t)=1$),  which allows to 
interpret them as probabilities and  (\ref{app:ClassicalNESS}) as a classical master equation for an auxiliary classical Markov process,  with renormalized rates $w_{\al \beta}$.  The stationary solution of the auxiliary Markov process (\ref{app:ClassicalNESS}),  i.e.  solution of
 \begin{align}
 \sum_{\be \neq \al} w_{\be \al} \nu_\be
  - \nu_\al \sum_{\be \neq \al} w_{\al \be}=0,
  \quad \al=1,2,\dots, 
  \label{eq:ClassicalNESSstationary}
\end{align}
where all $\nu_\al \geq 0$ due to Perron-Frobenius theorem,
yields the NESS spectrum $\{ \nu_\al \}\equiv \{ e^{- \tilde{E}_\al}/\tilde{Z}  \} $.  From (\ref{eq:ClassicalNESSstationary}) we can find the Zeno NESS  for 
various $\hat J$, $\vec{n}_l$ and $\vec{n}_r$.  
Numerically,  in all cases we  observe the  Kolmogorov
relation for the rates $w_{\alpha \beta}$
\begin{align}
  w_{\alpha\beta}w_{\beta\gamma}w_{\gamma\alpha}=w_{\alpha\ga}w_{\ga \beta}w_{\beta \alpha}. \label{eq:Kolmogorov}
\end{align}
Direct consequence of the Kolmogorov relation is the detailed balance property of the solution of (\ref{eq:ClassicalNESSstationary})
\begin{align}
 \nu_\al w_{\alpha\beta}=\nu_\be w_{\beta \alpha}\,. \label{eq:DetailedBalance}
\end{align}
\textit{Remark 2.}
The Kolmogorov relation (\ref{eq:Kolmogorov}) is only satisfied  for pure targeted boundary states,
i.e.  for boundary fields obeying (\ref{eq:h1}), (\ref{eq:hN}) with $|\vec n_l|= |\vec n_r|=1$.  
It is remarkable for our purpose since it gives a direct way to find the NESS spectrum $\nu_{\al}$, via (\ref{eq:DetailedBalance}), 
avoiding solution of the full linear system (\ref{eq:ClassicalNESSstationary}) which is in general of  exponential complexity (in system size $N$).  
Even though Eq. (\ref{eq:Kolmogorov}) is postulated,  in all specific cases its validity can be checked, or proved,  aposteriori.  Note that 
due to (\ref{eq:DetailedBalance}), the problem of finding  Zeno NESS eigenvalues $\nu_{\al}$ reduces to a problem of finding ratios of certain correlations 
of eigenstates of integrable model (\ref{eq:OpenXYZ}),  i.e. to a problem widely studied in the literature.  

Below we discuss several explicit examples, for which we find the NESS spectrum $\{\nu_{\al}\} = \{\exp(- \tilde E_\al)/\tilde{Z}\}$, and establish the dissipative dressing property (\ref{eq:dissDressing}):

(i) XXX and XXZ models with boundary qubits  targeted  in direction of $z$-axis;
(ii) XXZ model  with boundary targeting specific polarization in the  XY-plane;
(iii) XYZ model with boundary targeting in specific directions. 
In case (i) the corresponding effective Hamiltonian is $U(1)$-symmetric,  and in the case (ii) the $U(1)$-symmetry is broken by non-diagonal 
boundary fields.  Finally,  the case (iii) provides an evidence for the existence of
dissipatively dressed quasiparticles in the most general fully
anisotropic version of the Heisenberg exchange interaction, i.e., for
a XYZ spin chain, where $U(1)$ symmetry is broken already at the level
of the bulk Hamiltonian.
The original and modified dispersion relations, for all three cases (i), (ii), (iii) constitute our main results and are listed in the next section.

\section{Dissipative dressing of quasiparticle dispersion: XXX,XXZ, XYZ models with boundary driving}

\textit{XXX model with sink and source.---}
The dissipation-projected Hamiltonian with isotropic bulk exchange interaction and boundary fields along $z$-axis $h_1=- \si_1^z$,
$h_N = \si_N^z$,
\begin{align}
  H_D =
  \sum_{n=1}^{N-1} \vec{\si}_{n}\cdot \vec{\si}_{n+1} - \si_1^z + \si_N^z
  \label{openXXX}
\end{align} 
describes the effective dynamics (\ref{LMEeff}) of internal spins in an isotropic Heisenberg model of length $N+2$ where the first 
and the last spin are projected onto spin-down state $\ket{\downarrow}$ and spin-up state $\ket{\uparrow}$, respectively,
i.e.  $ \vec{n}_{l}=(0,0,-1)$,  $ \vec{n}_{r}=(0,0,1)$,  in accordance with (\ref{eq:h1}),  (\ref{eq:hN}).
The $U(1)$ symmetry renders the Hamiltonian (\ref{openXXX}) block-diagonalizable within blocks of fixed  magnetization. 
The eigenstates $\ket{\al}$ of $H_D$ belonging to the block with magnetization $N-2M$
 have energies (eigenvalues) $E_\al= N-1 + \sum_{j=1}^M \eps (u_{j,\al})$, where 
$u_{j,\al}$, $j=1,\dots,M$, are Bethe rapidities, satisfying the Bethe Ansatz relations
\begin{align}
  & \frac{u_j^{[-3]} }{ u_j^{[+3]} }
    \left(\frac{ u_j^{[+1]} }{ u_j^{[-1]} } \right)^{2N+1}
    = \prod_{\substack{k=1\\ k\neq j}}^M \prod_{\sigma=\pm1}
    \frac{(u_j+\sigma u_k)^{[+2]}}{(u_j+\sigma u_k)^{[-2]}}, 
    \label{eq:BAE}
\end{align} 
in which we defined 
\begin{align}
    u^{[q]}\equiv u + i q/2\,, 
\end{align}
and $\eps(u)$ is the standard dispersion relation, 
\begin{align}
&\eps (u) =  -\frac{2}{u^2+\frac{1}{4}}.
    \label{eq:eps(u)}
\end{align}

The NESS of the  spin chain with boundary dissipation Fig.\ref{FigSchema} has the form $\ket{\downarrow}\bra{\downarrow} \otimes
\rho_{\rm NESS} \otimes \ket{\uparrow}\bra{\uparrow} $ with $\rho_{\rm NESS}$ given by 
(\ref{eq:dissDressing}), $\tilde{E_\al} = \sum_{j=1}^M \tilde\eps (u_{j,\al}),$ and the 
modified, or dissipatively dressed,  dispersion relation
\begin{align}
&\tilde{E_\al} = \sum_{j=1}^M \tilde\eps (u_{j,\al}), \nonumber \\
&  \tilde{\eps}(u) =
  \log \left| \frac{u^2 +\frac{9}{4}}{u^2 +\frac{1}{4}} \right| 
  =\log  \left|1-\eps(u) \right|,
  \label{eq:tildeeps}
\end{align}
where $u_{j,\al}$ are given by the same BAE (\ref{eq:BAE}).

\textit{XXZ model with sink and source.---} Our next example is the XXZ model
\begin{align}
& H_{D} =\sum_{n=1}^{N-1}\vec{\si}_{n}\cdot  \hat {J} \vec{\si}_{n+1} +h_1 +h_N,
  \label{openXXZ}\\
& \hat {J} = {\rm diag}(1,1,\De), \quad h_1=-\si_1^z \De,  \quad  h_N= \si_N^z \De.
\end{align}
The Hamiltonian in Eq. (\ref{openXXZ})
gives an effective dynamics (\ref{LMEeff}) for XXZ spin chain with $z$-anisotropy $\De$ and  leftmost/rightmost  spins
projected onto  states $\ket{\downarrow}$/$\ket{\uparrow}$ , respectively,
i.e.  $ \vec{n}_{l}=(0,0,-1)$,  $ \vec{n}_{r}=(0,0,1)$,  see (\ref{eq:h1}),  (\ref{eq:hN}).  For $\De=1$, Eq. (\ref{openXXZ})
reduces to (\ref{openXXX}).  Like Eq. (\ref{openXXX}), the Hamiltonian (\ref{openXXZ}) possesses the $U(1)$ symmetry. 
The spectrum of coherent model (\ref{openXXZ}) is given by $E_\al= (N-1)\De +  \sum_{j=1}^M \eps (u_{j,\al})$, 
where $\{u_{j,\al}\}$  satisfy Eq.  
(\ref{eq:BAE}) with the replacement
\begin{align}
u^{[k]} \equiv \sinh(u+ik\ga/2)\,,
\label{trig}
\end{align}
where $\cos \ga=\De$,  see  pioneering
paper of 
Sklyanin~\cite{1988Sklyanin}.  The NESS 
spectrum of the dissipation-driven model Fig.~\ref{FigSchema} 
is given by $\{ \nu_\al \propto \exp(- \tilde E_\al) \} $,   $\tilde {E}_\al=  \sum_{j=1}^M \tilde\eps (u_{j,\al})$.

The original and dissipatively dressed dispersions are, respectively, 
\begin{align}
\eps(u) &= \frac{-2 \sin^2 \ga}  {\sinh(u+\frac{i\ga}{2})\sinh(u-\frac{i\ga}{2})}.\label{eq:eps(u)Trig}\\
\tilde \eps(u)&= \log \left|\frac{\sinh(u+\frac{3i\ga}{2}) \ \sinh(u-\frac{3i\ga}{2})}{\sinh(u+\frac{i\ga}{2}) \  \sinh(u-\frac{i\ga}{2})} \right|\no\\
&=\log|1-\eps(u)\De|.
 \label{eq:eps(u)DissTrig}
\end{align}
Setting $\ga \rightarrow \de$,
$u \rightarrow \de u $ and letting $\de \rightarrow 0$ one recovers
the result for the isotropic Heisenberg model (\ref{eq:tildeeps}).

\textit{XXZ model with chiral invariant subspace.---} An XXZ Hamiltonian (\ref{openXXZ}) with 
non-diagonal boundary fields 
\begin{align}
&h_1=\si_1^x,\quad h_N=\si_N^x  \cos \varphi(M)  + \si_N^y  \sin \varphi(M),\label{openXXZchiral}\\
&\varphi(M)=(N+1-2M)\ga,\quad \cos\gamma=\De,\label{eq:PhiM}
\end{align} 
gives an effective dynamics (\ref{LMEeff}) for XXZ spin chain with anisotropy $\De$ and  leftmost/rightmost  spins
projected onto  fully polarized states in XY-plane $\vec{n}_l=(1,0,0)$  $\vec{n}_r=(\cos \varphi(M),\sin \varphi(M),0)$. 
It was shown in Refs. \cite{2021PhantomLong,2022DissCooling} that for
integer values $M=0,1,\dots,N+1 $ the Hamiltonian (\ref{openXXZchiral}) has a chiral
invariant subspace $G_M$.  $G_M$ is
 spanned by pieces of spin helices of the 
period $2\pi/\ga$,  and of the same helicity sign, and its dimension is
$d_M= \binom{N+1}{M}$.   

The Gibbs state restricted to the invariant subspace $G_M$ has the
form~\eqref{eq:rhoGibbs} with $\al=1,\dots,d_M$ and $j=1,\dots,M$.
The integer $M$ counts the number of kinks and is analogous to the number of spin down arrows in the $U(1)$-invariant case 
(\ref{openXXZ}): each state $\ket{\al}$ is parametrized via $M$ Bethe rapidities $u_{j,\al}$, satisfying Bethe Ansatz equations:
\begin{align}
&\left[\frac{\sinh(u_j+\frac{i\ga}{2})}{\sinh(u_j-\frac{i\ga}{2})}\right]^{2N+2}\left[\frac{\cosh(u_j -\frac{i\ga}{2})}{\cosh(u_j +\frac{i\ga}{2})}\right]^2\nonumber\\
&=\prod_{\sigma=\pm1}\prod_{k\neq j}^M\frac{\sinh(u_j+\sigma u_k+i\ga)}{\sinh(u_j+\sigma u_k-i\ga)},\quad j=1,\ldots,M.\label{BAE;XXZ1}
\end{align}
The energy levels of the coherent model (\ref{openXXZ}),  (\ref{openXXZchiral})
and of the dissipatively constrained model $E_\al$,  $\nu_\al \propto \exp(-\tilde E_\al )$ are given by 
usual  $E_\al= (N+1)\De+ \sum_{j=1}^M \eps(u_{j,\al})$, $\tilde E_\al = \sum_{j=1}^M \tilde \eps(u_{j,\al})$,
with original and dissipatively dressed dispersions 
\begin{align}
\eps(u) 
  &= \frac{-2 \sin^2 \ga}  {\sinh(u+\frac{i\ga}{2})\sinh(u-\frac{i\ga}{2}) },
  \label{eq:eps(u)Trig}\\
  \tilde \eps(u)
    &= 2 \log \left|
    \frac {\cosh \left(u+\frac{i\ga}{2} \right)
    \cosh \left(u-\frac{i\ga}{2} \right)}
    {\sinh \left(u+\frac{i\ga}{2} \right)\sinh \left(u-\frac{i\ga}{2}\right)}
    \right|\no\\
&=2\log\left|\frac{\De\,\eps(v)}{2(1-\De^2)}-1\right|. \label{eq:XXZdressingChiral}
\end{align}

Note that the original dispersion $\eps(u)$ is the same as in the $U(1)$ invariant XXZ case even though the nature of the eigenstates $\ket{\al}$
 is here 
completely different.  Note also that after a standard substitution $e^{ip} = \frac{\sinh(u+\frac{i\ga}{2})}{\sinh(u-\frac{i\ga}{2})}$
the dispersion relation $\eps(u)$ acquires more familiar form  $\eps(p)=4 (\cos p - \De)$.

All  eigenstates within $G_M$  are chiral,  since they are a linear superposition of spin helix pieces with the same helicity
(\ref{eq:XXZdressingChiral}).  The ``chiral quasiparticles" do not carry fixed
magnetization as in the  $U(1)$ case (\ref{openXXZ}), but rather form
domain walls, or kinks, on top of a chiral ``background'', see
\cite{2021PhantomBetheAnsatz} for more details.  
Equation (\ref{eq:XXZdressingChiral}) is proved in Section \ref{sec:M1Chiral}, while its generalization 
for $M>1$  is a conjecture based on numerics, see Appendix~\ref{appendix:XXZchiral}.

\textit{XYZ model with chiral invariant subspace.---} Finally we
consider  fully anisotropic XYZ spin chain
\begin{align}
  &H_{D} =
  \sum_{n=1}^{N-1} \vec{\si}_{n} \hat {J} \vec{\si}_{n+1}
  +h_1(\vec{n}_l)+h_N(\vec{n}_r),
  \label{openXYZchiral}\\
&\hat{J} = \mathrm{diag}(J_x,\,J_y,\,J_z),\no
\end{align}
where $h_1$ and $h_N$ are given by (\ref{eq:h1}), (\ref{eq:hN}).  
$J_\alpha$ can be parametrized in terms of two complex parameters $\eta,\tau$
and the Jacobi $\theta$-functions as
$
  \left\{J_x,J_y,J_z\right\}=\left\{ \frac{\bell{4}(\eta)}{\bell{4}(0)},  
  \, \frac{\bell{3}(\eta)}{\bell{3}(0)}, \   \frac{\bell{2}(\eta)}{\bell{2}(0)}\right\}.$
Following Ref. \cite{WatsonBook}, we use the shorthand notation
$\bell{\alpha}(u)\equiv\vartheta_\alpha(\pi u|e^{i\pi\tau})$,
$\ell{\alpha}(u)\equiv\vartheta_\alpha(\pi u|e^{2i\pi\tau})$.
Real $\eta$ and purely imaginary $\tau$ with $\imag(\tau)>0$ give real $J_\al$.

As explained in Section \ref{sec:EffectiveDynamics}, (\ref{openXYZchiral}) describes an effective dynamics (\ref{LMEeff}) for XYZ spin chain with  boundary spins dissipatively projected onto fully polarized states $\vec{n}_l$,$\vec{n}_r$.
If is convenient to parametrize unit vectors $\vec{n}_l,\vec{n}_r$  by two complex parameters $u_l=x_l + i y_l,\ u_r =
x_r + i y_r$
as \cite{2022XYZ}

\begin{align}
  &n_{ l,r}^x = -\frac{ \bell{2}(i y_{ l,r})}{\bell{3}(i y_{ l,r})} \   \frac{ \bell{1}(x_{ l,r})}{\bell{4}(x_{ l,r})}, \nonumber\\
  &n_{ l,r}^y = - i \frac{ \bell{1}(i y_{ l,r})}{\bell{3}(i y_{ l,r})} \   \frac{ \bell{2}(x_{ l,r})}{\bell{4}(x_{ l,r})},  \label{eq:nvec-parametrization}\\  
  &n_{ l,r}^z =- \frac{ \bell{4}(i y_{ l,r})}{\bell{3}(i y_{ l,r})} \   \frac{ \bell{3}(x_{ l,r})}{\bell{4}(x_{ l,r})}. \nonumber
\end{align}
We focus on eigenstates of (\ref{openXYZchiral}) within a chiral invariant subspace $G_M$ with dimension $\binom{N+1}{M}$,  which appears for $u_r=u_l+(N+1-2M)\eta$
and integer $M\in[0,\,N+1]$ \cite{2022XYZ}.  This chiral subspace is an elliptic extension of the spin-helix based chiral subspace 
in the previous example.  Each state $\ket{\al}$ within the invariant subspace
is parametrized via $M$ Bethe rapidities $\{u_{j,\al}\}$, which satisfy the following BAE \cite{Yang2006}
\begin{align}
&\left[\frac{\bell{1}(u_j+\frac{\eta}{2})}{\bell{1}(u_j-\frac{\eta}{2})}\right]^{2N+2}\frac{\bell{3}(u_j +iy_l-\frac{\eta}{2})}{\bell{3}(u_j -iy_l+\frac{\eta}{2})}\frac{\bell{4}(u_j +x_l-\frac{\eta}{2})}{\bell{4}(u_j-x_l+\frac{\eta}{2})}\nonumber\\
&\times\frac{\bell{3}(u_j-iy_r-\frac{\eta}{2})}{\bell{3}(u_j+iy_r+\frac{\eta}{2})}\frac{\bell{4}(u_j-x_r-\frac{\eta}{2})}{\bell{4}(u_j+x_r+\frac{\eta}{2})}\nonumber\\
&
=\prod_{\sigma=\pm1} \prod_{k\neq j}^M\frac{\bell{1}(u_j+\sigma u_k+\eta)}{\bell{1}(u_j+\sigma u_k-\eta)},\quad  j=1,\ldots,M.\label{BAE;XYZ;M}
\end{align}
The energy of the model can be expressed in terms of Bethe rapidities $\{u_{j,\al}\}$ as
\begin{align}
E_\al =&\, \sum_{j=1}^M \eps(u_{j,\al})+(N+1)g(\eta )\no\\
&\,+g\left(\tfrac{\tau }{2}+x_l\right)-g\left(\tfrac{\tau }{2}+x_r\right), \\
g(u) =&\, \frac{\bell{1}(\eta)\bell{1}'(u)}{\bell{1}'(0) \bell{1}(u)},\quad \bell{1}'(u)=\frac{\partial \bell{1}(u)}{\partial u},
\end{align}
where $\eps(u)$ is the original dispersion relation
\begin{align}
  \begin{split}
    &\eps(u) = 2 \left[g(u-\tfrac{\eta }{2})-g(u+\tfrac{\eta }{2})\right].
  \end{split}
  \label{eq:XYZChiralDispersion}
\end{align}

On the basis of numerical studies (see Appendix \ref{appendix:XYZchiral}), we 
conjecture that dissipative dressing effect takes place also here,  i.e.  
the $\rho_{\rm NESS}= \tilde{Z}^{-1}\sum_{\al} \exp(- \tilde{E}_\al) \ket{\al} \bra{\al}$
with $ \tilde{E}_\al=\sum_{j=1}^M \tilde \eps(u_{j,\al})$,  where 
\begin{align}
  \begin{split}
    &\tilde \eps(u) =
      2\log \left| \frac{Q\left(  \frac{1-\tau}{2} +
      i \IM u_l, u  \right) }{Q(0, u)} \right|,
    \\
    &Q(x,u) =
      \th_1 \left(x-u+\tfrac{\eta}{2} \right)
      \th_1\left(x+u+\tfrac{\eta}{2} \right),
  \end{split}
  \label{eq:XYZdressingChiral}
\end{align} 

By letting $i \IM u_l = \tau/2$ and $\tau \rightarrow +i \infty$, one
recovers from (\ref{eq:XYZdressingChiral}) the XXZ limit (\ref{eq:XXZdressingChiral}) with $\gamma=\pi\eta$.
More details are given in Appendix~\ref{appendix:XYZchiral}.

\section{XXX/XXZ model with sink and source: derivation of dissipative dressing relations
}
The derivation of the expressions for dissipatively dressed dispersion appears rather technical as any problem having to do with
finding
correlation functions of interacting many-body systems.   Here we show how to derive the simplest of our results, 
Eq.  (\ref{eq:tildeeps}) for the dissipatively dressed dispersion of isotropic Heisenberg model with sink a source,  in  a pedagogical way.

The eigenstates of 
\begin{align}
H_D=\sum_{n=1}^{N-1} \vec{\si}_{n}\cdot \vec{\si}_{n+1} - \si_1^z + \si_N^z,\quad \label{app:openXXX}
\end{align} 
can be separated into blocks with fixed total magnetization due to $U(1)$ symmetry. Since $\rho_{\rm NESS}$ commutes 
with (\ref{app:openXXX}),  $H$ and $\rho_{\rm NESS}$ can be diagonalized simultaneously.  
The simplest eigenstate of $H$
is a  state with maximal possible magnetization $\ket{0} \equiv \ket {\uparrow  \uparrow \ldots \uparrow} $.
In the block with $M=1,2,\ldots$ spins down the eigenstates have the form
\begin{align}
&\ket{\al}=\sum_{n=1}^N \ga_n(u_\alpha)\,\si_n^{-} \ket{0},\label{eq:psiM=1}\\
&\ket{\beta} = \sum_{n_1,n_2=1}^N \ga_{n_1,n_2}(u_{1,\beta},u_{2,\beta})\  \si_{n_1}^{-} \si_{n_2}^{-} \ket{0},\label{eq:psiM=2}\\
&\ldots \nonumber
\end{align}
Assuming non-degeneracy of the spectrum of $H$,
$\rho_{\rm NESS}$ has the same eigenstates and can be written as
\begin{align}
&\rho_{\rm NESS} = \nu_0 \ket{0} \bra{0} + \sum_{\al=1}^N \nu_\al^{(1)}  \ket{\al} \bra{\al}+ \ldots 
\end{align}
 where $ \ldots$ denote contributions from blocks with $M>1$.
From Eqs. (\ref{eq:gL}), (\ref{eq:gR}), we find $g_l= \si_1^{-}$,  $g_r= \si_N^{+}$.

 According to detailed balance relations (\ref{eq:DetailedBalance})  (that can be proved aposteriori),  
$\nu_\al^{(1)}/\nu_0 =  w_{0 \al} /  w_{ \al  0}$.   From (\ref{app:rates}) we find
\begin{align}
\frac{w_{0\al}}{w_{\al 0}} &= \frac{ |\bra{\al} \si_1^{-} \ket{0}|^2 +  |\bra{\al} \si_N^{+} \ket{0}|^2}
{ |\bra{0} \si_1^{-} \ket{\al}|^2 +  |\bra{0} \si_N^{+} \ket{\al}|^2}  \\
&=\frac{|\bra{\al} \si_1^{-} \ket{0}|^2}{|\bra{0} \si_N^{+} \ket{\al}|^2}= \left| \frac{ \ga_1}{\ga_N} \right|^2
\end{align} 

In order  to find the ratio $\ga_1/\ga_N$: we   use the algebraic Bethe Ansatz method. 
Within the method,  the eigenvectors of $H$ are produced by repetitive  action of a ``creation" operator $\BB(u)$ on the vacuum state $\ket{0}$.  Each application of $\BB(u)$ lowers the total
magnetization by $1$, so e.g.  eigenstates from (\ref{eq:psiM=1}), (\ref{eq:psiM=2}) are produced 
as $\ket{\al} = \BB(u_{1,\al}) \ket{0}$, $\ket{\be} = \BB(u_{1,\be})\BB(u_{2,\be}) \ket{0}$.
Here $u_{j,\al}, u_{j,\be}$ are Bethe roots (solutions of (\ref{eq:BAE}) )  for $M=1$ and $M=2$ respectively.  

Let us first introduce the double-row monodromy matrix 
\begin{align}
\mathcal{T}_0(u)=T_0(u)\mathcal{K}_0(u)\bar T_0(u),\label{monodromy}
\end{align}
where subscript 0 denotes the two-dimensional auxiliary space. The one-row monodromy matrices $\tau(u),\bar\tau(u)$ and the $K$-matrix in Eq. (\ref{monodromy}) are defined as 
 \begin{align}
&T(u)=R_{0,N}(u)\ldots R_{0,1}(u)=\left(
\begin{array}{cc}
A(u) & B(u) \\
C(u) & D(u) \\
\end{array}\right),\\
&\bar T(u)=R_{0,1}(u)\ldots  R_{0,N}(u)=\left(
\begin{array}{cc}
\bar A(u) & \bar B(u) \\
\bar C(u) & \bar D(u) \\
\end{array}\right),\\
&R_{0,n}(u)= 
 \left( 
\begin{array}{cc}
u I +\frac{i}{2} \si_n^z& i \si_n^{-}\\[4pt]
 i \si_n^{+}& u I -\frac{i}{2} \si_n^z
\end{array}
\right),
\end{align}
and 

\begin{align}
\mathcal{K}(u)={\rm diag}\Big\{u^{[-3]},\,\,-u^{[1]}\Big\}.
\end{align}
The ``eigenstates" creation operator  $\BB(u)$ is an element of $\mathcal{T}(u)$: 
\begin{align}
\BB(u)=[\mathcal{T}(u)]^1_2=u^{[-3]}A(u)\bar B(u)-u^{[1]}B(u)\bar{D}(u).
\end{align}
The following relation
\begin{align}
T_1(u)R_{1,2}(2u\!-\!\tfrac{i}{2})\bar T_2(u)\!=\!\bar T_2(u)R_{1,2}(2u\!-\!\tfrac{i}{2})T_1(u),
\end{align}
leads to
\begin{align}
A(u)\bar B(u)=\frac{2u-i}{2u}\bar{B}(u)A(u)-\frac{i}{2u}B(u)\bar{D}(u).
\end{align}
Then, we derive
\begin{align}
\BB(u_{1,\al})
=&\frac{u^{[-3]}u^{[-1]}}{u}\bar{B}(u)A(u)\no\\
-&\frac{u^{[3]}u^{[-1]}}{u}B(u)\bar{D}(u)\,.
\end{align}
From the explicit form of the one-row monodromy matrices,
we obtain
 \begin{align}
&A(u)\ket{0}=\Big(u^{[1]}\Big)^{N}\ket{0}, \\
& \bar D(u)\ket{0}=\Big(u^{[-1]}\Big)^{N}\ket{0}, \\
&\bar B(u)\ket{0} =i \sum_{m=1}^N F_m(u)\,  \sigma_m^-\ket{0}, \\
&B(u)\ket{0} =i \sum_{m=1}^N F_{N-m+1}(u) \si_m^- \ket{0}, \\
&F_m(u)= \Big(u^{[1]}\Big)^{m-1} \Big(u^{[-1]}\Big)^{N-m}.
\end{align}
The Bethe state $\BB(u) \ket{0}$ can be expanded as
\begin{align}
&\BB(u) \ket{0}=i\frac{u^{[-3]}u^{[-1]}}{u}\Big(u^{[1]}\Big)^{N}\sum_{m=1}^N F_m(u)\,  \sigma_m^-\ket{0}\no\\
&\quad-i\frac{u^{[3]}u^{[-1]}}{u}\Big(u^{[-1]}\Big)^N\sum_{m=1}^N F_{N-m+1}(u) \si_m^- \ket{0}.\label{BS;XXX}
\end{align}
With the help of Eq. (\ref{BS;XXX}), it can be derived that 
\begin{align}
\gamma_1(u)&=2\Big(u^{[-1]}\Big)^N\Big(u^{[1]}\Big)^{N-1},\\
\gamma_N(u)&=\frac{i u^{[-3]} u^{[-1]}\Big(u^{[1]}\Big)^{2N-1}-iu^{[3]}\Big(u^{[-1]}\Big)^{2N}}{u}\no\\
&\overset{(\ref{eq:BAE})}{=}2\Big(u^{[-1]}\Big)^{2N}u^{[3]}\Big(u^{[1]}\Big)^{-2}.
\end{align}
One can thus obtain the ratio 
\begin{align}
\frac{\gamma_1^2(u)}{\gamma_N^2(u)}=\left(\frac{u^{[1]}}{u^{[-1]}}\right)^{2N}\left(\frac{u^{[1]}}{u^{[3]}}\right)^2\overset{(\ref{eq:BAE})}{=}\frac{u^{[1]}u^{[-1]}}{u^{[3]}u^{[-3]}},\label{eq:ratioNuMresult}
\end{align}
leading to (\ref{eq:tildeeps}) for $M=1$.

Proceeding iteratively for $M=2,3,\ldots$ one can verify the general validity of (\ref{eq:tildeeps}).

The  effects of a dissipative dressing on the NESS spectrum can be seen already in the one-particle 
sector $M=1$. 
 Notably,  the bare (original) quasiparticle dispersion $\eps(u)$ has a singularity at $u = \pm i/2$, 
while dissipatively dressed quasiparticle dispersion $\tilde{\eps}(u)$ (\ref{eq:ratioNuMresult}) 
retains the  singularity at $u = \pm i/2$ and acquires  an extra singularity at 
$u= \pm 3i/2$.  In the following we show that some solutions of BAE lie
exponentially close to the extra singularity $u= \pm 3i/2$,   drastically modifying the NESS spectrum $\tilde E_\al$ with respect 
to the $H$ spectrum $E_\al$.

\begin{figure}
  \centering \includegraphics[width=0.45\textwidth]{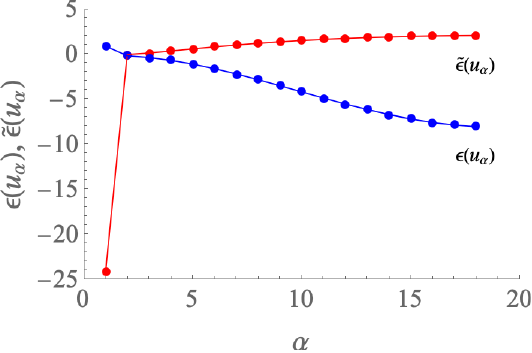}
  \includegraphics[width=0.45\textwidth]{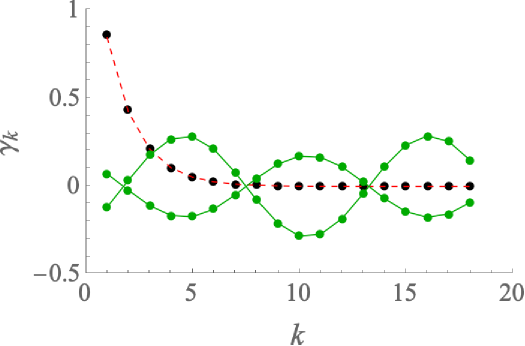}
  \caption{ Top panel: quasiparticle energies $\eps(u_\al)$ (blue
    joined points) and their dissipative dressing $\tilde \eps(u_\al)$ (red joined points) in the
    XXX model with $N=18$ spin, in the block with one magnon $M=1$.
    The state $\al=1$ is a localized Bethe state with
    $u_1\simeq 3i/2 +i e^{-24.5}$.  Bottom
    panel: coefficients $\ga_k$ of the normalized localized Bethe state
    $\ket{\al=1} = \sum_{k=1}^N \ga_k\ \si_k^{-} \ket{0}$ (black empty
    circles). The dashed red line is the fit
    $\ga_k = 1.7 \times 2^{-k}$.  The green joined points are the
    coefficients $\RE \ga_k$ and $\IM \ga_k$ for the plain-wave like Bethe
    state with $u_4\approx 1.78139$.   }
  \label{FigM1}
\end{figure}

The sector $M=1$, contains
$N$ Bethe eigenstates $\ket{\al}$ parametrized by
the solutions $u_\al$, $\al=1,\ldots N$, of the BAE (\ref{eq:BAE}).
Among the $N$ BAE solutions there are always $N-1$ real
solutions, say $u_2, \ldots, u_N$, and one boundary localized
imaginary solution $u_1$,  lying exponentially close to the extra
singularity due to dissipative dressing, namely,
$u_1=i(3 /2 + O( 2^{-N}))$, see the top panel of Fig.~\ref{FigM1}.
Explicitly, from (\ref{eq:BAE}) we find
$u_1-3 i/2 \equiv \de = 3 i 2^{-2N-1} (1+ O(N\de))$. The corresponding
dressed energy $\tilde\eps(u_1)$ is drastically renormalized by the
singularity, acquiring a negative amplitude linearly growing with
system size, $\tilde\eps(u_1) \approx -2(N+1)\log 2 +\log 9$, see (\ref{res:XXXrenormalizationBoundStates}). 
On the other hand,  for plane wave type (real $u_j$) BAE solutions,
the dressed and original energies are comparable.  As a result,
$e^{-\tilde \eps(u_1)} \gg e^{-\tilde \eps(u_\al)}$ for $\al>1$, and
in the NESS, the boundary localized Bethe eigenstate $\ket{\alpha=1}$ comes
with an exponentially large (in system size $N$) relative weight in
the sum~(\ref{eq:NESS}) with respect to the plane wave $M=1$ eigenstates
$\ket{\alpha}$, see  Fig.~\ref{FigM1}. In sectors with several quasiparticles $M>1$ we observe similar phenomenon   which leads to  an overall subextensive scaling of the 
von-Neumann entropy of the NESS (\ref{eq:NESS}) with the system size, see \cite{2024NESSspecArxiv}.

\textit{Trigonometric case.~---}
We pass to the trigonometric XXZ case (\ref{openXXZ}) by redefining the functions $u^{[k]}$ as (\ref{trig}). The dispersion relation of the quasiparticles has form 
\begin{align}
&\eps(u) = -\frac{2 \sin^2 \gamma}{u^{[1]}u^{[-1]}} 
\label{app:eps(u)Trig}
\end{align}
and the XXZ eigenvalues are
\begin{align}
&E = \sum_{j=1}^M \eps(u_j) + (N-1)\cos \gamma,
\label{app:ETrig}
\end{align}
where $\{u_j\}$ are solutions of BAE (\ref{eq:BAE}),  with redefinition
(\ref{trig}).
The one-particle wave function is given by the  expression (\ref{BS;XXX})
with the replacements  (\ref{trig}) and $ i \rightarrow i\sin \gamma$.

Repeating the calculations, we find 
\begin{align}
\frac{\nu_\al^{(1)}}{\nu_0} &= \frac {u^{[1]} u^{[-1]} }{u^{[3]} u^{[-3]} }  \label{eq:ratioNuM1Trig}
\end{align}
leading to (\ref{eq:eps(u)DissTrig}) for $M=1$.

Proceeding analogously for $M=2,3,\ldots$ one can verify the validity of (\ref{eq:eps(u)DissTrig}), and consequently,
of (\ref{eq:tildeeps}),  iteratively.  
In \cite{2024NESSspecArxiv} we obtain the result (\ref{eq:eps(u)DissTrig}) in an alternative way, 
expressing the operators $g_l,g_r$ in terms of  $\BB(u)$ and using the commutation relation 
between the monodromy matrix elements.

Let us analyze the BAE solutions for $M=1$ case in more detail.
The corresponding Bethe states can be rewritten in terms of the quasi-momentum $p$ as
\begin{align}
&\sum_{n=1}^N\mathcal{A}_n(p)\,\sigma_n^-\ket{0},\\
& \mathcal{A}_n(p)= (e^{i p}-2\Delta)e^{inp}-\left(1-2\Delta  e^{i p}\right)e^{i(1-n)p},\label{BS;M1}
\end{align}
where $p$ satisfies
\begin{align}
e^{i(2N+1) p}(e^{ip}-2 \De)-(1-2\De e^{ip})=0.\label{BAE;M1}
\end{align}
Most solutions for $p$ of (\ref{BAE;M1}) are real. However, in some cases, Eq. (\ref{BAE;M1}) has an imaginary solution. 
Indeed, let us define a function 
\begin{align}
\mathcal{W}(x)=x^{(2N+1)}(x-2 \De)-(1-2\De x). \label{eq:W(x)}
\end{align}
One can easily check that 
\begin{align}
&\mathcal{W}(1)=0,\quad \mathcal{W}(\tfrac{1}{2\De})=(1-4 \Delta ^2) (2\De)^{-2N-2},\no\\
&\mathcal{W}(2\De)=4\De^2-1,\quad \mathcal{W}'(1)=2 + 2N (1- 2\De).
\end{align}
Once
\begin{align}
\De>\frac{1}{2}, \quad N>\frac{1}{2\De-1}\label{imag;eq}
\end{align}
we have
\begin{align}
\mathcal{W}(\tfrac{1}{2\De})<0,\quad \mathcal{W}(1)=0,\quad \mathcal{W}'(1)<0. 
\end{align}
Therefore, $\mathcal{W}(x)=0$ has a solution in the interval $(\frac{1}{2\De},1)$ under the condition (\ref{imag;eq}).
And, from (\ref{eq:W(x)}), the root  approaches $\frac{1}{2\De}$ as $N$ increases.
Since $\mathcal{W}(e^{i p})=0$ is exactly the BAE (\ref{BAE;M1}),  this implies an imaginary root $p$ with $e^{ip}\approx\frac{1}{2\De}$ (or equivalently, $u\approx-\frac{3i\gamma}{2}$).
When $N$ is large enough, this imaginary $p$ will gives a localized state with 
\begin{align}
\mathcal{A}_n(p)\approx \left(\frac{1}{2\De}-2\Delta\right)\left(\frac{1}{2\De}\right)^n.\label{Localized;State}
\end{align} 

The dressed energy in terms of $p$ is 
\begin{align}
&\quad\log|e^{-ip}(e^{ip}-2\De)(1-2\De e^{ip})|\no\\
&\overset{(\ref{BAE;M1})}{=}\log|e^{2iNp}(e^{ip}-2\De)^2|.
\end{align}
For the imaginary solution $e^{ip}\approx\frac{1}{2\De}$ for large $N$,  the dressed energy $\tilde{\eps}(p)$ becomes
\begin{align}
\tilde{\eps} &\approx\log|(2\De)^{-2N-2}(1-4\De^2)^2| \no\\
&=-2(N+1)\log(2\De)+2\log(4\De^2-1). \no
\end{align}  
In particular, for $\De=1 $  we obtain
\begin{align}
&\tilde{\eps}\approx-2(N+1)\log 2+\log 9. \label{res:XXXrenormalizationBoundStates}
\end{align}  

\section{XXZ model with chiral invariant subspace: derivation of dissipatively dressed dispersion (\ref{eq:XXZdressingChiral})}
\label{sec:M1Chiral}

Like in the previous section, we first prove (\ref{eq:XXZdressingChiral}) for the single quasiparticle $M=1$ sector. 

Let us introduce the following family of chiral  states:
\begin{align}
  \begin{split}
    &\bra{\Phi(n_1,\ldots,n_M)}
      =\exp\left[i\ga\sum_k n_k\right]\bigotimes_{l_1=1}^{n_1}\phi(l_1)\\
&\bigotimes_{l_2=n_1+1}^{n_2}\phi(l_2-2)
      \cdots \bigotimes_{l_{M+1}=n_M+1}^{N}\phi(l_{M+1}-2M),
    \\
    &\phi(n)=\frac{1}{\sqrt{2}}(1,\,e^{-i n\gamma}).
  \end{split}
  \label{Basis;XXZ}
\end{align} 
Then, the set
$$\bra{\Phi(n_1,\ldots,n_M)},\qquad 0\leq n_1<n_2\cdots<n_M\leq N,$$
forms an invariant subspace of the Hamiltonian (\ref{openXXZ}),(\ref{openXXZchiral})  \cite{2021PhantomLong}.
A chiral analog of coordinate  Bethe Ansatz \cite{2021PhantomBetheAnsatz} then leads to 
the consistency conditions  which have the form of standard BAE (\ref{BAE;XXZ1}).  
   We aim at finding NESS eigenvalues,  corresponding to the 
  invariant subspace eigenstates $\ket{\al}$ of (\ref{openXXZ}),(\ref{openXXZchiral}).  
According to (\ref{eq:DetailedBalance}), the ratio of the NESS eigenvalues for the associated dissipative 
problem (Fig.~\ref{FigSchema}) is given by 

\begin{align}
&\frac{\nu_\al}{\nu_\be} =  \frac{w_{\be\al}}{w_{\al \be}} = \frac{ |\bra{\al} g_l \ket{\be}|^2 +  |\bra{\al}g_r \ket{0}|^2}
{ |\bra{\be} g_l \ket{\al}|^2 +  |\bra{\be} g_r \ket{\al}|^2}, \label{eq:NuRatioChiral}
\end{align} 
with operators $g_{l,r}$ computed from (\ref{eq:gL}),(\ref{eq:gR}):
\begin{align}
\begin{aligned}
&g_l=G_L \otimes I^{\otimes_{N-1}},\quad g_r=I^{\otimes_{N-1}} \otimes G_R,\\
& G_{L,R}=\left(
\begin{array}{cc}
\cos\gamma & -e^{-i\vfi_{l,r}} \\
e^{i\vfi_{l,r}} & -\cos \gamma \\
\end{array}\right),\\
&\vfi_l=0,\,\,\vfi_r=\vfi(M).
\end{aligned}
\end{align}
Let us  calculate  (\ref{eq:NuRatioChiral}) for $M=1$, where the Bethe states have the form  
\cite{2021PhantomLong}
\begin{align}
  &\bra{\al} = \sum_{n=0}^N \bra{\Phi(n)} f_n(p^{(\alpha)}),\label{M1;XXZ}\\
  &f_n(p) = e^{i\chi + i n p} + e^{-i\chi - i n p},\quad e^{2i \chi(p)} =  \frac{\De-e^{ip}}{e^{-ip}-\De}.\label{eq:Chi}
\end{align} 
Here, the ``quasimomentum" $p$ and the rapidity $u$ from (\ref{BAE;XXZ1}) are related by standard relation
 \begin{align}
  &e^{i p }=\frac{\sinh(u+\frac{i\ga}{2})}{\sinh(u-\frac{i\ga}{2})}. \label{eq:P(U)}
\end{align}
In terms of $p$, BAE  (\ref{BAE;XXZ1}) for $M=1$ acquire a simple form
\begin{align}
  &e^{2i N p}e^{4i\chi}=1.\label{BAE1}
\end{align} 
All solutions $p_\al$ of Eq.~(\ref{BAE1}) are real, meaning that also $f_n(p_\al)$ are all real. 

To compute the expressions $\bra{\al} g_l \ket{\be}$ in (\ref{eq:NuRatioChiral}) we note
the following property:
\begin{align}
\begin{split}
&g_l\ket{\Phi(n)}=\ka(2\de_{n,0}-1)\ket{\Phi(n)}, \\
&g_r \ket{\Phi(n)} = \ka(1- 2\de_{n,N})  \ket{\Phi(n)}, 
\end{split} \label{gLRprop}\\
&\ka = i \sin \ga.\no
\end{align} 
While the eigenstates $\ket{\al}$ in (\ref{M1;XXZ}) are orthonormal,  the chiral  states 
$\ket{\Phi(n)}$ are not,  
\begin{align} \braket{\Phi(n)}{\Phi(m)} = \De^{|n-m|}. \label{PhiOverlap}
\end{align} 
For further calculations we set 
\begin{align} 
p\equiv p_{\alpha},  \ p'\equiv p_\beta, \quad   \chi\equiv\chi(p_\alpha),  \  \chi'\equiv\chi(p_\beta)\label{def:ChiralNotations}
\end{align} 
for notational simplicity. 
Using (\ref{gLRprop}), (\ref{PhiOverlap}), $\braket{\al}{\be}=0$ and $f_n^*=f_n$ 
we obtain
\begin{align}
&\bra{\al} g_l \ket{\be} =2 \ka f_0(p') \sum_{n=0}^N f_n(p)\De^n, \label{rates1}\\
&\bra{\al} g_r \ket{\be} = - 2 \ka f_N(p') \sum_{n=0}^N f_{N-n}(p)\De^n.\label{rates2}
\end{align} 
From (\ref{BAE1}), we readily find
\begin{align}
  &f_{N-n}(p) = \pm f_n(p).  \label{CN}
\end{align} 
It follows that
\begin{align}
  & |\bra{\al} g_l \ket{\be}|=  |\bra{\al} g_r \ket{\be}|, \label{R1}
\end{align} 
and consequently
\begin{align}
  &\frac{\nu_\al}{\nu_\be} =  \frac  {|\bra{\al} g_l \ket{\be}|^2}{ |\bra{\be} g_l \ket{\al}|^2}.
    \label{LambdaRatio}
\end{align} 
Performing  summation in (\ref{rates1}) we obtain
\begin{align}
  \bra{\al} g_l \ket{\be} &= 2 \ka f_0(p')\sum_{n=0}^N f_n(p) \De^n
                            \no\\
                          &=2 \ka f_0(p') \left( 
                            e^{i\chi} \frac{1-z_{+}^{N+1} }{1-z_{+}}  + e^{-i\chi} \frac{1-z_{-}^{N+1} }{1-z_{-}}
                            \right),
  \no\\
                          z_{\pm} &= e^{\pm i p } \De. \no
\end{align} 
 Remarkably,  due to BAE (\ref{BAE1}),  the $N$-dependence in the above vanishes:
\begin{align}
&  e^{i\chi} \frac{z_{+}^{N+1} }{1-z_{+}}  + e^{-i\chi} \frac{z_{-}^{N+1} }{1-z_{-}}\no\\
  &=z_-^{N+1}e^{-i\chi} \left(
    \frac{e^{2 i Np + 2 i \chi} e^{2 i p}} {1-z_{+}} + \frac{1}{1-z_{-}}
    \right)\no\\
  &= z_-^{N+1}e^{-i\chi} 
    \left(
    \frac{e^{- 2 i \chi} e^{2 i p}} {1-z_{+}} + \frac{1}{1-z_{-}}
    \right)\no\\
  &=z_-^{N+1}e^{-i\chi} 
    \left(
    e^{-ip} \frac{1-z_{+}}{\De -e^{ip}} \frac{ e^{2 i p}} {1-z_{+}} + \frac{1}{1-z_{-}}
    \right)\no\\
  &=0.
\end{align} 
So we obtain
\begin{align}
  &\frac{\bra{\al} g_l \ket{\be} }{2 \ka } = f_0(p') \left( 
    \frac{e^{i\chi} }{1-z_{+}}  + \frac{e^{-i\chi} }{1-z_{-}}
    \right).
    \label{ss2}
\end{align} 
From (\ref{eq:Chi}) we have
\begin{align}
  &e^{i\chi} (e^{-ip} - \De) = e^{-i\chi} (\De - e^{ip}),
    \no
\end{align} 
and for the real part 
\begin{align}
  &\cos(\chi-p) = \De \cos \chi.
    \label{cos(Q-p)}
\end{align} 
Noting $f_0(p') = 2 \cos \chi'$ and using (\ref{cos(Q-p)}),  after some straightforward algebra we obtain
\begin{align}
  &\bra{\al} g_l \ket{\be} = \frac{8\ka (\De^{-1}-\De)  \cos \chi' \cos \chi}{ \De+ \De^{-1}  - 2\cos p}.
\end{align} 
The expression for $\bra{\be}g_l \ket{\al}$ is obtained from the
above by substitutions $p\leftrightarrow p'$,
$\chi\leftrightarrow \chi'$,  yielding 
\begin{align}
  &\frac{\bra{\al}g_l \ket{\be}} {\bra{\be}g_l \ket{\al}}  = \frac{\De+\De^{-1}- 2\cos p' }{\De+\De^{-1}  - 2\cos p }.
    \label{ss3}
\end{align} 
Notably,   the Kolmogorov relation  (\ref{eq:Kolmogorov}) follows from Eq. (\ref{ss3}), justifying aposteriori the 
assumption (\ref{eq:NuRatioChiral}).

Restoring the original notations (\ref{def:ChiralNotations}) we get
\begin{align}
  &\frac{\nu_\al} {\nu_\be}=
 \frac{|\De+\De^{-1}  - 2\cos p_\be |^2}{|\De+\De^{-1}  - 2\cos p_\al|^2 }.
    \label{ResM1}
\end{align} 
Finally,  using  $\log\frac{\nu_\be} {\nu_\al} = \tilde{E}_\al - \tilde{E}_\be=
 \tilde{\eps} (u_\al)  - \tilde{\eps} (u_\be)$,  and  (\ref{eq:P(U)}),  the Eq.~(\ref{ResM1}) reduces to
(\ref{eq:XXZdressingChiral}).

\begin{figure}
  \centering \includegraphics[width=0.45\textwidth]{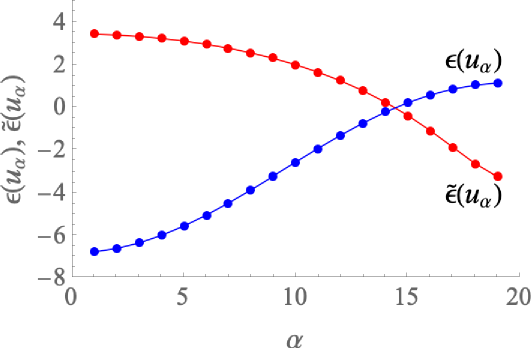}
  \caption{ Quasiparticle energies $\eps(u_\al)$ (blue
    joined points) and dissipatively dressed ones  $\tilde \eps(u_\al)$ (red joined points) in the
    XXZ model with chiral invariant subspace, in the block with one kink $M=1$.   Parameters:
 $N=18$, $\De=0.7$.  There are no localized states: all the amplitudes  are plain-wave 
like,  data not shown. }
  \label{FigM1chiral}
\end{figure}

Fig.~\ref{FigM1chiral} shows dissipatively dressed and original energies for the multiplet 
from chiral invariant subspace with  $M=1$.  
We see that the the dissipative dressing reverses the order of the contributions in the respective NESS with respect to the Gibbs state,
(for $\eps(u_\al)<\eps(u_\be)$, $\tilde \eps(u_\al)>\tilde \eps(u_\be)$).
However, unlike in the XXZ model with diagonal   boundary fields,  we do not see  drastic renormalization
effects for some eigenstates like  in  Fig.~\ref{FigM1}. This happens because  
 all  Bethe roots  $u_j$
are now real,  and therefore  cannot approach singularities of $\tilde \eps(u)$,
lying on the imaginary axis,  as it happened 
in the ``sink and source" case (Fig.~\ref{FigM1}).

Even though there are no principal difficulties to extend the above calculations for larger $M$
 using the chiral Bethe Ansatz method    \cite{2021PhantomBetheAnsatz},
the analytic expressions become
difficult to handle  because of  the non-orthogonality of the chiral basis vectors (\ref{Basis;XXZ}). 
For $M>1$ the validity of (\ref{eq:XXZdressingChiral}) is thus checked numerically, see Appendix \ref{appendix:XXZchiral}.

\section{Discussion}

In summary, we have found and described a surprising effect of
  dissipative dressing of quasiparticle dispersion relation, in integrable spin chains attached to stronly dissipative baths.
So far we have been able to find a  rigorous proof only for the  $U(1)$-symmetric 
``sink and source" case (\ref{eq:eps(u)DissTrig}),
which is valid for all eigenstates, see the companion paper \cite{2024NESSspecArxiv}.  
However, the  effect of dissipative dressing is definitely present in other integrable systems as discussed in the present paper, including fully anisotropic  XYZ model, as suggested by firm numerical 
evidence.   
 For all the cases,  we find the dissipatively dressed dispersion relation (energy of quasiparticle) of 
  the form
  \begin{align}
    &\tilde \eps(u) \sim \log (\eps(u)/f(u)), \label{ExtraSingularityProperty}
  \end{align} 
  where $u$ is a Bethe rapidity, $\eps(u)$ is the original
  quasiparticle dispersion, and $f(u)$ contains an additional
  singularity.  While for the XXZ model the property
  (\ref{ExtraSingularityProperty}) is obvious by comparison of
  (\ref{eq:eps(u)DissTrig}) and (\ref{eq:XXZdressingChiral}) with
  (\ref{eq:eps(u)Trig}), for the XYZ model
  (\ref{ExtraSingularityProperty}) seems not obviously true.  However, also for
  the XYZ case, our numerical data confirm the validity of
  (\ref{ExtraSingularityProperty}) where $f(u)$ depends on extra
  complex parameter (to be reported elsewhere).

Using the dissipatively dressed dispersion relation as a key input we have been able to
diagonalize the steady state density operators of
boundary dissipatively driven integrable quantum spin chains in the
limit of large dissipation, alias the Zeno regime,
by employing the Bethe Ansatz of a related 
dissipation-projected Hamiltonian. We have thus discovered a simple but surprising phenomenon of ``dissipative dressing" of quasiparticle energies in integrable coherent systems exposed to boundary dissipation.

Our results should have applications in state engineering and
dissipative state preparation. Moreover, we expect analogous emergent
integrability of the steady state density operators in the discrete-time case of an
integrable Floquet XXX/XXZ/XYZ circuits, where boundary dissipation can
be conveniently implemented by the so-called reset channel~\cite{2024Abanin,PopkovProsen2025}.

\bigskip
\begin{acknowledgements}
  V.P. and T.P. acknowledge support by ERC Advanced grant
  No.~101096208 -- QUEST, and Research Program P1-0402 and Grant N1-0368 of Slovenian
  Research and Innovation Agency (ARIS). V.P. is also supported by
  Deutsche Forschungsgemeinschaft through DFG project
  KL645/20-2. X.Z. acknowledges financial support from the National
  Natural Science Foundation of China (No. 12204519).
\end{acknowledgements}


 \bibliographystyle{apsrev4-1} \bibliography{NessSpecRef}

\clearpage

\onecolumngrid

\appendix

\setcounter{page}{1}
\setcounter{equation}{0}
\setcounter{figure}{0}

\renewcommand{\theequation}{\textsc{A}-\arabic{equation}}
\renewcommand{\thefigure}{\textsc{A}-\arabic{figure}}

\begin{center}
  {\bfseries {\em APPENDICES }\\
  }
\end{center}

This Appendix part contains three sections.  In Appendix ~\ref{app:ZenoBrickwall}  we sketch a derivation of the 
Lindblad Master equation starting from brickwall qubit circuit.  Appendices~\ref{appendix:XXZchiral} and
\ref{appendix:XYZchiral} contain some details and Tables for  XXZ and XYZ model with 
chiral invariant subspaces.

\section{Lindblad Master equation evolution as a limit of a brickwall unitary circuit 
}
\label{app:ZenoBrickwall}
Here we describe a  brickwall unitary circuit leading to the Lindblad Master equation 
Eq.(\ref{LME}) with an XXX Hamiltonian in the continuous time (Trotter) limit.

\begin{figure}
  \centering
 \includegraphics[width=0.45\columnwidth,clip]{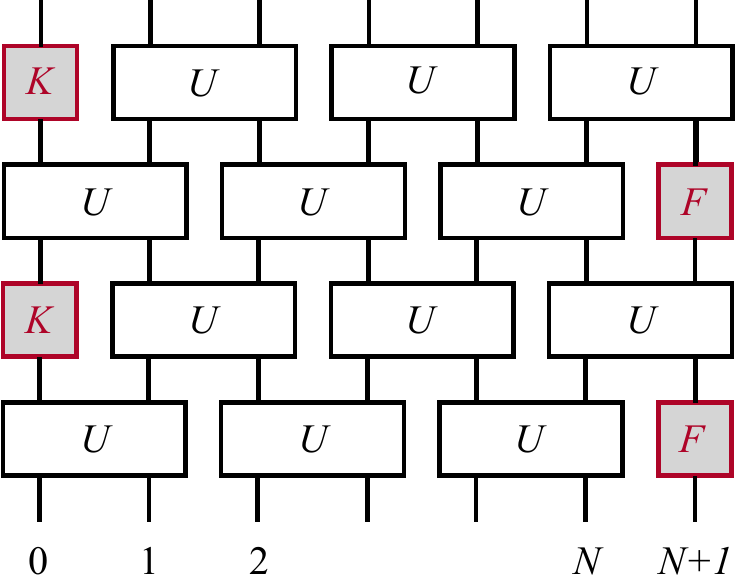}
 \hspace{0.05\columnwidth}
 \includegraphics[width=0.315\columnwidth,clip]{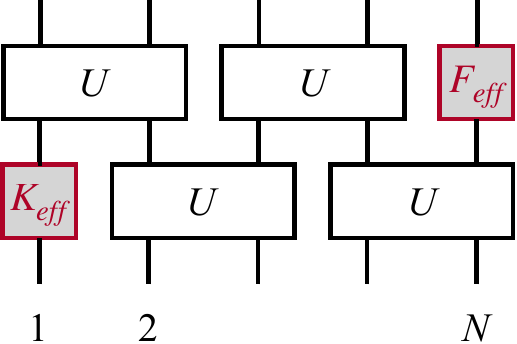}
  \caption{ Brickwall unitary circuit (BUC), original (left) and effective (right).  
Left panel shows original BUC with reset gates $K$ and $F$ at the edges.  Two-body interaction $U$ is given by
(\ref{Ugate}).
Right Panel
shows effective BUC,  where the first site  and the last site are traced out,  giving rise to 
effective Krauss operators $K_{eff}$ and $F_{eff}$.}
  \label{FigBrickwall}
\end{figure}

It is described by a two step discrete time protocol,  schematically shown in Fig.~\ref{FigBrickwall}.
Assuming $N$ being odd,  the discrete evolution has the following form
\begin{align}
&{\cal M}={\cal M}_{\rm odd} {\cal M}_{\rm even} \\
&{\cal M}_{\rm even}[\rho] =\sum_{j=1}^2 F_j(\eps)
U_{01} U_{23} \ldots U_{N-1,N} \  \rho \ U_{01}^\dagger  U_{23}^\dagger  \ldots U_{N-1,N}^\dagger F_j(\eps)^\dagger\\
&{\cal M}_{\rm odd}[\rho] =\sum_{j=1}^2 K_j(\eps)
U_{12}  U_{34} \ldots U_{N,N+1}\   \rho \  U_{12}^\dagger  U_{34}^\dagger \ldots U_{N,N+1}^\dagger K_j(\eps)^\dagger\\
&U_{n,n+1}= \frac{1- i \tau P_{n,n+1}}{1- i \tau}, \quad P_{n,n+1} = \frac12 \left( I + \vec{\si}_n \cdot \vec{\si}_{n+1} \right)
\label{Ugate}
\end{align} 
where $P_{n,n+1}$ is a permutation,  
and the Krauss gates $K_j(\eps)$,  $\tau$ is a real number,  $F_j(\eps)$,  $0\leq \eps<1$ act on spins $0$ and $N+1$ respectively,  see
Fig.~\ref{FigBrickwall}.  
Explicitly the Krauss gates are given by 
\begin{align}
& F_1(\eps) = \sqrt{1-\eps} \ \si_{N+1}^{+},\quad F_2(\eps) =\left(
\begin{array}{cc}
1  &0 \\
0& \sqrt{\eps} 
\end{array}
\right)  \label{Fgate}\\
& K_1(\eps) = \sqrt{1-\eps}\  \si_{0}^{-},\quad K_2(\eps) =\left(
\begin{array}{cc}
\sqrt{\eps}  &0 \\
0& 1
\end{array}
\right) \label{Kgate}
\end{align} 

The Krauss gate $K(\eps)[\cdot] =\sum_{j=1}^2 K_j [\cdot] K_j^\dagger $ is a linear operation on a $2\times 2$  matrices.  Its  eigenstates $\psi_k$ and eigenvalues 
$\la_k$ are
 $\psi_0= \ket{\uparrow}\bra{\uparrow}$,  $\psi_1= \si^z$, $\psi_2= \si^+$, $\psi_3= \si^{-}$, 
and $\la_0=1, \la_1 =\eps $,  $\la_3 =\la_4 = \sqrt{\eps} $. 
 On the other hand,  the exponentiated Lindblad operator  $e^{ \Ga t \,{\cal D}_{\si^{-}}} [\cdot]$ with ${\cal D}_L$
given by (\ref{LindbladOperator}) has exactly the same eigenvectors $\{ \psi_k \}$ and the eigenvalues $\mu_0=1, \mu_1=e^{-\Ga t}$,
$\mu_2=\mu_3=e^{-\Ga t/2}  $.  We thus obtain,  that an application of a Krauss gate  $n$ times 
 \begin{align}
K^n(\eps)[\cdot] = e^{ \Ga t \,{\cal D}_{\si^{-}}} [\cdot],  \quad \mbox {if}\ \   \eps =  e^{-\Ga t/n}.
\end{align}
Analogously,  
 \begin{align}
F^n(\eps)[\cdot] = e^{ \Ga t \,{\cal D}_{\si^{+}}} [\cdot],  \quad \mbox {if}\ \   \eps =  e^{-\Ga t/n}.
\end{align}

The boundary driven Lindblad equation (\ref{LME})  is obtained in the  limit  
\begin{align}
&\tau=  t/n \ll 1,\\ 
&U \approx I - i \tau P\\
&{\cal M}[\rho(t)] \equiv  \rho(t+\tau)
\end{align}
so that $\lim_{\tau\rightarrow 0} ({\cal M}[\rho(t)] - \rho(t))/\tau = \frac{d \rho}{d t}$ after a straightforward algebra
 leads to the differential equation
 (\ref{LME}). We thus can view the brickwall unitary circuit as a Trotter discretization scheme of the Lindblad Master 
equation (\ref{LME}).

Quantum Zeno limit corresponds to $\eps=0$ in ( \ref{Fgate}),  ( \ref{Kgate}),  when the $K$ and $F$ gates 
are projecting the spins  on pure states $\ket{\downarrow} $ and $\ket{\uparrow} $ respectively,
i.e.  $K[\cdot],F[\cdot]$ become the so-called reset gates:
\begin{align}
&\sum_{j=1}^2 F_j(0) \rho_{N+1}  F_j(0)^\dagger = \ket{\uparrow}\bra{\uparrow},  \quad \rho_{N+1}=\mathrm{tr}_{0,1,\ldots N} \rho\\
&\sum_{j=1}^2 K_j(0) \rho_{0}  K_j(0)^\dagger = \ket{\downarrow}\bra{\downarrow},  \quad \rho_{0}=\mathrm{tr}_{1,\ldots N,N+1} \rho.
\end{align} 
Indeed, an application of the $K$ and $F$ gates each elementary time step $\tau$ can be  viewed as an effective realization of a protocol of repeated interactions,  leading to quantum Zeno regime.  As a result, 
the  quantum chain splits effectively in three parts: spin $``0"$( always in the state  $\ket{\downarrow}\bra{\downarrow}$)
 $+$ internal spins $``1,2,\ldots,N"$ $+$ spin $``N+1"$( always in the state  $\ket{\uparrow}\bra{\uparrow}$).
The bulk interaction affecting the internal spins $1,2,\ldots,N$
gets renormalized, resulting in the application of the effective single spin boundary channels: 
\begin{align}
&K_{\rm eff}(\rho) = \mathrm{tr}_1 \left(U_{12} (\ket{\downarrow}\bra{\downarrow} \otimes \rho) U^\dagger_{12}\right),\\
&F_{\rm eff}(\rho) = \mathrm{tr}_2 \left(U_{12} (\rho\otimes \ket{\uparrow}\bra{\uparrow})U^\dagger_{12}\right),
\end{align}
acting on  spins $1$, $N$, respectively: see Fig.~\ref{FigBrickwall}, lower Panel.    
It can be shown (to be reported  elsewhere) that 
the quantum Zeno NESS (the fixed point of the effective time evolution of the brickwall circuit for $\tau\rightarrow 0$)
commutes with the effective Hamiltonian  (\ref{openXXX}).

The above scheme  is straightforwardly extendable 
on all other cases considered in our manuscript.

 
\section{Numerical evidence showing consistency of  Eq.~(\ref{eq:XXZdressingChiral})}
\label{appendix:XXZchiral}

For cases when $M > 1$, we currently lack full analytic proof of the (\ref{eq:XXZdressingChiral}).
Nevertheless, this equation can be verified numerically for $M=2$ and larger $M$. The ratio ${\nu_\beta}/{\nu_\alpha}$ obtained
from Eq. (\ref{eq:XXZdressingChiral}) consistently matches other approaches. 
Some explicit examples are shown in
Tables \ref{Tab1} and \ref{Tab2}, for $M=2$ and $M=3$, where we employ the following BAE equivalent to Eq. (\ref{BAE;XXZ1}) 
\begin{align} e^{2i(N+1)p_j}\left[\frac{\De-e^{ip_j}}{1-\De e^{ip_j}}\right]^2=\prod_{\sigma=\pm}\prod_{k\neq j}^M\frac{1- 2 \De e^{ip_j}+  e^{ip_j}e^{i\sigma p_k}} {1- 2 \De e^{i\sigma p_k}+ e^{ip_j}e^{i\sigma p_k}},\quad j=1,\ldots,M,\label{BAE;XXZ}
\end{align}
and $\eps(p)=4(\cos p-\De)$.

\begin{table}[htbp]
  \begin{tabular}{|c|c|c|c|}
    \hline
    $p_1^{(\alpha)}$ & $p_2^{(\alpha)}$ & $E$ & $\log(\nu_\alpha/\nu_1)$\\
    \hline 
    2.6041 & 2.0574 & $-$6.6398 & 0 \\
    1.5086 & 2.6122 & $-$4.5374 & 0.8315 \\
    1.5169 & 2.0803 & $-$3.0687 & 1.2440 \\
    0.9512 & 2.6300 & $-$2.4983 & 2.2403 \\
    0.4126 & 2.6596 & $-$1.2134 & 4.4264 \\
    2.1192 & 0.9537 & $-$1.1040 & 2.6273 \\
    0.4052 & 2.1843 & 0.0399 & 4.8052 \\
    0.9620 & 1.6082 & 0.8043 & 3.3183 \\
    1.7257 & 0.3885 & 1.7515 & 5.4714 \\
    1.2864+0.8655$i$ & 1.2864$-$0.8655$i$ & 1.8059 & 1.5936 \\
    1.0386+0.2488$i$ & 1.0386$-$0.2488$i$ & 2.8522 & 4.2450 \\
    0.3566 & 1.2991 & 3.4884 & 6.4596 \\
    0.7405 & 0.8791 & 4.1706 & 5.9499 \\
    0.9129 & 0.3134 & 4.9178 & 7.7571 \\
    0.2771 & 0.5625 & 5.8978 & 9.2974 \\
    \hline 
  \end{tabular}
  \caption{Numerical solutions of BAE (\ref{BAE;XXZ}) and the
    entanglement NESS spectrum. Here $N=5,M=2$ and
    $\De=\frac23$.}\label{Tab1}
\end{table}

\begin{table}[htbp]
  \begin{tabular}{|c|c|c|c|c|}
    \hline 
    $p_1^{(\alpha)}$ & $p_2^{(\alpha)}$ & $p_3^{(\alpha)}$ & $E$ & $\log(\nu_\al/\nu_1)$\\
    \hline 
    2.1710 & 2.6609 & 1.6539 & $-$9.4713 & 0 \\
    1.1143 & 2.1955 & 2.6721 & $-$7.4769 & 1.1630 \\
    2.2510 & 0.5090 & 3.5861 & $-$5.9674 & 3.3356 \\
    1.7101 & 1.1263 & 2.676 & $-$5.7429 & 1.7502 \\
    2.2068 & 1.7148 & 1.1306 & $-$4.5789 & 2.0594 \\
    2.7009 & 1.8138 & 0.4928 & $-$4.3896 & 3.9236 \\
    2.6694 & 1.2008+0.6111$i$ & 1.2008$-$0.6111$i$ & $-$3.4460 & 1.7685 \\
    2.2621 & 0.4860 & 1.8187 & $-$3.3285 & 4.2366 \\
    0.4550 & 1.3969 & 2.7080 & $-$2.6780 & 4.8481 \\
    2.1861 & 1.2128$-$0.6428$i$ & 1.2128+0.6428$i$ & $-$2.2398 & 1.9781 \\
    3.5856 & 0.9363$-$0.1577$i$ & 0.9363+0.1577$i$ & $-$2.1442 & 4.0653 \\
    1.4017 & 2.2764 & 0.4486 & $-$1.6501 & 5.1467 \\
    0.3894 & 1.0133 & 3.5639 & $-$1.1651 & 6.1495 \\
    2.2529 & 0.9381$-$0.1605$i$ & 0.9381+0.1605$i$ & $-$1.0633 & 4.3462 \\
    1.6596 & 1.2492$-$0.7461$i$ & 1.2492+0.7461$i$ & $-$0.4221 & 2.2348 \\
    0.3843 & 1.0153 & 2.2998 & $-$0.1800 & 6.4309 \\
    0.4343 & 1.8503 & 1.4124 & $-$0.1772 & 5.6766 \\
    2.7341 & 0.6441 & 0.3290 & $-$0.0218 & 7.7711 \\
    1.8046 & 0.9432+0.1685$i$ & 0.9432$-$0.1685$i$ & 0.5048 & 4.8354 \\
    0.3250 & 0.6419 & 2.3311 & 0.9047 & 8.0426 \\
    1.0197 & 1.8869 & 0.3738 & 1.2413 & 6.9218 \\
    1.0909 & 1.3074$-$0.9403$i$ & 1.3074+0.9403$i$ & 1.5866 & 2.6908 \\
    1.9386 & 0.3170 & 0.6368 & 2.2449 & 8.5127 \\
    1.3068 & 0.9753$-$0.2139$i$ & 0.9753+0.2139$i$ & 2.3009 & 5.5199 \\
    6.6372 & 1.4846 & 1.0309 & 2.8193 & 7.6528 \\
    0.5572 & 1.3522+1.1229$i$ & 1.3522$-$1.1229$i$ & 3.0105 & 3.8879 \\
    0.4728 & 1.1858$-$0.5741$i$ & 1.1858+0.5741$i$ & 3.7407 & 6.5804 \\
    0.3033 & 0.6254 & 1.5671 & 3.7418 & 9.2056 \\
    1.4664 & 0.6801+0.4587$i$ & 0.6801$-$0.4587$i$ & 3.9690 & 6.6551 \\
    0.3363 & 1.0639$-$0.3143$i$ & 1.0639+0.3143$i$ & 4.5199 & 8.4020 \\
    0.6724 & 0.8847$-$0.0706$i$ & 0.8847+0.0706$i$ & 4.8769 & 8.2448 \\
    0.5961 & 1.2358 & 0.2830 & 5.1327 & 10.1424 \\
    0.2880 & 0.9049$-$0.1041$i$ & 0.9049+0.1041$i$ & 5.4709 & 9.8548 \\
    0.2598 & 0.9604 & 0.5368 & 6.2627 & 11.2949 \\
    0.4815 & 0.7035 & 0.2388 & 7.1487 & 12.5699 \\
    \hline
  \end{tabular}
  \caption{Numerical solutions of BAE (\ref{BAE;XXZ}) and the
    entanglement NESS spectrum. Here $N=6$, $M=3$, and
    $\De =\frac23 $.}\label{Tab2}
\end{table}

\section{Numerical evidence showing consistency of  Eq.~(\ref{eq:XYZdressingChiral})}
\label{appendix:XYZchiral}

For XYZ model,  we are able to guess the analytic expression for the dissipatively dressed dispersion
(\ref{eq:XYZdressingChiral}).
Some technical details followed by an explicit examples are given  below.

Define the following state
\begin{align}
  &\ket{\Psi(n_1,\ldots,n_M)}=\bigotimes_{l_1=1}^{n_1}\psi(l_1)\bigotimes_{l_2=n_1+1}^{n_2}\psi(l_2-2)\cdots \bigotimes_{l_{M+1}=n_M+1}^{N}\psi(l_{M+1}-2M),\label{Basis;XYZ}\\
  &\psi(x)=\binom{\ell{1}(\varepsilon+x\eta)}{-\ell{4}(\varepsilon+x\eta)},\qquad \varepsilon=u_l=x_l+iy_l.
\end{align}
Similarly to the XXZ case with XY plane boundary targeting, the set
$$\ket{\Psi(n_1,\ldots,n_M)},\qquad 0\leq n_1<n_2\cdots<n_M\leq N,$$
forms a chiral invariant subspace of  
\begin{align}
  H_{D} =
  \sum_{n=1}^{N-1} \vec{\si}_{n} \hat {J} \vec{\si}_{n+1}
  +h_1(\vec{n}_l)+h_N(\vec{n}_r),
  \label{app:XYZchiral}
\end{align}
if special values of $\vec{n}_l$, $\vec{n}_r$ are chosen, see Eq. (\ref{eq:nvec-parametrization}) in the main text. 
 
One can use the chiral states (\ref{Basis;XYZ})
to expand the Bethe state inside the  invariant subspace
and the corresponding expansion
coefficients depend on the Bethe roots $\{u_1,\ldots,u_M\}$ in
(\ref{BAE;XYZ;M}) \cite{2022XYZ}.

Under the dynamics of the XYZ Hamiltonian, the Zeno NESS  has
reduced rank $d_M= \binom {N+1}{M}$, namely,
\begin{align}
  &\rho_{\rm NESS}= \sum_{\al=1}^{d_M}\nu_\al \ket{\al} \bra {\al}.  \label{app:ZenoXYZ}
\end{align}
where $\ket{\al}$ are eigenstates of $H_D$ (\ref{app:XYZchiral})
 belonging to the chiral invariant subspace.

The operators $g_{l,r}$ can be written in terms of elliptic functions as  follows:
\begin{align}
&g_l=G_L \otimes I^{\otimes_{N-1}},\quad g_r=I^{\otimes_{N-1}} \otimes G_R,\ \no \\
&G_{L,R}=\left(
\begin{array}{cc}
\frac{\bell{2}(\eta ) \sqrt{\bell{1}(x_{l,r}-i y_{l,r}) \bell{1}(x_{l,r}+i y_{ l,r})}}{\bell{4}(x_{l,r}) \bell{3}(i y_{l,r})} & -\frac{\bell{1}(x_{ l,r}-i y_{l,r}) \left[\bell{4}(\eta ) \bell{3}(x_{l,r}+i y_{l,r})-\bell{3}(\eta ) \bell{4}(x_{l,r}+i y_{l,r})\right]}{\bell{4}(x_{l,r}) \bell{3}(i y_{l,r}) \sqrt{\bell{1}(x_{l,r}-i y_{l,r}) \bell{1}(x_{l,r}+i y_{l,r})}}\\
-\frac{\bell{1}(x_{ l,r}-i y_{ l,r}) \left[\bell{3}(\eta ) \bell{4}(x_{ l,r}+i y_{ l,r})+\bell{4}(\eta ) \bell{3}(x_{ l,r}+i y_{ l,r})\right]}{\bell{4}(x_{ l,r}) \bell{3}(i y_{ l,r}) \sqrt{\bell{1}(x_{ l,r}-i y_{ l,r}) \bell{1}(x_{ l,r}+i y_{ l,r})}} & -\frac{\bell{2}(\eta ) \sqrt{\bell{1}(x_{ l,r}-i y_{ l,r}) \bell{1}(x_{ l,r}+i y_{ l,r})}}{\bell{4}(x_{ l,r}) \bell{3}(i y_{ l,r})}
\end{array}
\right).
\end{align}
 Similarly to the XXZ case, one can prove that all chiral states $\ket{\Psi(n_1,\ldots,n_M)}$
are eigenstates of $g_l$ and  $g_r$:
\begin{align}
&g_l \ket{\Psi(s,n,m, \ldots)} = (1-2\delta_{s,0})\ka_l  \ket{\Psi(s,n,m, \ldots )},\\
&g_r \ket{\Psi(n,m, \ldots,s)} = (2\delta_{s,N}-1)\ka_r  \ket{\Psi(n,m, \ldots,s)},\\
&\kappa_{l,r}=\frac{\sqrt{\bell{1}(x_{l,r}-i y_{l,r})} }{\sqrt{\bell{1}(x_{l,r}+i y_{l,r})} }\frac{\bell{1}(\eta )\bell{2}(x_{l,r}+i y_{l,r})}{\bell{4}(x_{l,r}) \bell{3}(i y_{l,r})}.
\end{align}

Define the function
\begin{align}
Q(x,\{y_k\})=\prod_{k=1}^M\bell{1}(x-y_k+\tfrac{\eta}{2})\bell{1}(x+y_k+\tfrac{\eta}{2}).\label{Q;XYZ}
\end{align}
\begin{hyp}
The spectrum $\nu_\al$ of the Zeno NESS  (\ref{app:ZenoXYZ}) is given by 
\begin{align}
\frac{\nu_\be} {\nu_\al}
=\left|\frac{Q(\frac{1-\tau}{2}+iy_l,\{u_k^{(\alpha)}\})\,Q(0,\{u_k^{(\beta)}\})}{Q(0,\{u_k^{(\alpha)}\})\,Q(\frac{1-\tau}{2}+iy_l,\{u_k^{(\beta)}\})}\right|^2,\label{Hypothesis;XYZ}
\end{align}
where $\{u_j^{(\alpha)}\}$ and $\{u_j^{(\beta)}\}$ are the Bethe roots corresponding to $\ket{\al}$ and $\ket{\beta}$ respectively.
\end{hyp}

Analytic calculations for the XYZ model are extremely complicated due to the involvement of elliptic functions. Therefore, we resort to numerical calculations to verify our hypothesis.

Based on numerical results for the case $M=1$, we find the following simple and elegant expression:
\begin{align}
\frac{\nu_\be} {\nu_\al}=\frac{w_{\al \be}}{w_{\be\al}}=\left|\frac{Q(b,u^{(\alpha)})}{Q(a,u^{(\alpha)})}\frac{Q(a,u^{(\beta)})}{Q(b,u^{(\beta)})}\right|^2.
\end{align}
where $a$ and $b$ are two fixed system-dependent parameters.
Some numerical data for the values of $a$ and $b$ is
\begin{align}
\begin{array}{|c|c|c|c|c|c|}
\hline 
\tau & \eta &  x_l & y_l & a & b \\ 
\hline 
0.35i & 0.47 & 0.16  & 0.05 & 0 & 0.5-0.125i \\
\hline 
0.35i & 0.47 & 0.71  & 0.05 & 0 & 0.5-0.125i \\
\hline 
0.35i & 0.47 & 0.16  & 0.67 & 0 & 0.5+0.495i \\
\hline 
0.35i & 0.47 & 0.16  & 0.49 & 0 & 0.5+0.315i \\
\hline 
0.35i & 0.47 & 0.16  & 0 & 0 & 0.5-0.175i \\
\hline 
0.35i & 0.55 & 0.16  & 0.05 & 0 & 0.5-0.125i \\
\hline 
0.49i & 0.47 & 0.16  & 0.75 & 0 & 0.5+0.505i \\
\hline 
\end{array}
\end{align}
On the base of numerics  we conclude that
 $$a=0,\qquad  b=\frac{1-\tau}{2}+iy_l.$$
For larger $M$, numerical results (see e.g.  Tabs. \ref{Tab;XYZ1} and \ref{Tab;XYZ2}
for an illustration) also indicate that
\begin{align}
\frac{\nu_\be} {\nu_\al}&=\frac{w_{\al \be}}{w_{\be\al}}=\left|\frac{Q(\frac{1-\tau}{2}+iy_l,\{u_k^{(\alpha)}\})\,Q(0,\{u_k^{(\beta)}\})}{Q(0,\{u_k^{(\alpha)}\})\,Q(\frac{1-\tau}{2}+iy_l,\{u_k^{(\beta)}\})}\right|^2,
\end{align}
where $Q$ is given by Eq.~(\ref{Q;XYZ}). Equation (\ref{Hypothesis;XYZ}) notably results in the dressed energy given by Eq.~(\ref{eq:XYZdressingChiral}).

\begin{table}[htbp]
\begin{tabular}{|c|c|c|c|}
\hline 
$u_1^{(\alpha)}$ & $u_2^{(\alpha)}$ & $E$ & $\nu_\alpha/\nu_1$\\
\hline 
 0.0532$i$ & 0.2393$i$ & $-$17.5382 & 1\\
 0.0532$i$ & 0.8950+0.1750$i$ & $-$16.0715 & 1.0877\\
 0.1106$i$ & 0.1050+0.1750$i$ & $-$15.5292 & 1.1215\\
 0.2972$i$ & 0.3050+0.1750$i$ & $-$3.2127 & 2.3593\\
 0.1100$i$ & 0.6950+0.1750$i$ & $-$2.6720 & 2.4323\\
 0.1050+0.1750$i$ & 0.3050+0.1750$i$ & $-$1.1998 & 2.6465 \\
 0.3032$i$ & 0.5000+0.2451$i$ & 0.7863 & 2.8787 \\
 0.2668$i$ & 0.5000+0.2638$i$ & 1.2244 & 2.9460\\
 0.7650+0.1643$i$ & 0.7650+0.1857$i$ & 1.2589 & 3.0816\\
 0.1192$i$ & 0.5000+0.0645$i$ & 1.6803 & 3.0172\\
 0.1050+0.1750$i$ & 0.5000+0.1174$i$ & 2.7827 & 3.2291\\
 0.5000+0.0581$i$ & 0.1050+0.1750$i$ & 3.1085 & 3.2729 \\
 0.5000+0.2359$i$ & 0.3050+0.1750$i$ & 15.6611 & 7.0102\\
 0.5000+0.0559$i$ & 0.6950+0.1750$i$ & 15.9818 & 7.1036 \\
 0.5000+0.1141$i$ & 0.5000+0.2941$i$ & 19.9809 & 8.6736\\
\hline
\end{tabular}
\caption{Numerical solutions of BAE (\ref{BAE;XYZ;M}) and the entanglement NESS spectrum. Here $\eta =0.47$, $N=5$, $M=2$, $\tau =0.35 i$, $\{x_l,y_l\}=\{0.13,0.21\}$.}\label{Tab;XYZ1}
\end{table}
\begin{table}[htbp]
\begin{tabular}{|c|c|c|c|c|}
\hline
$u_1^{(\alpha)}$ & $u_2^{(\alpha)}$ & $u_3^{(\alpha)}$ & $E$ & $\nu_\alpha/\nu_{10}$ \\
\hline
 0.1050+0.1750$i$ & 0.1107$i$ & 0.2967$i$ & $-$26.4413 & 0.3315 \\
 0.0520$i$ & 0.2412$i$ & 0.3650+0.1750$i$ & $-$11.1247 & 0.8185 \\
 0.8954+0.1750$i$ & 0.3651+0.1750$i$ & 0.0520$i$ & $-$9.6473 & 0.8907 \\
 0.6351+0.1750$i$ & 0.8946+0.1750$i$ & 0.2980$i$ & $-$9.6357 & 0.8914 \\
 0.1088$i$ & 0.1046+0.1750$i$ & 0.6348+0.1750$i$ & $-$9.1113 & 0.9181 \\
 0.1088$i$ & 0.6352+0.1750$i$ & 0.1054+0.1750$i$ & $-$9.1005 & 0.9187 \\
 0.8950+0.1750$i$ & 0.0470$i$ & 0.5000+0.1076$i$ & $-$8.1303 & 0.9539 \\
 0.5000+0.0900$i$ & 0.0846$i$ & 0.1050+0.1750$i$ & $-$7.6879 & 0.9765 \\
 0.8950+0.1750$i$ & 0.6594+0.1750$i$ & 0.1294+0.1750$i$ & $-$7.6318 & 1.0058 \\
 0.5000+0.0664$i$ & 0.8950+0.1750$i$ & 0.1194$i$ & $-$7.2318 & 1 \\
 0.5000+0.1037$i$ & 0.0458$i$ & 0.3650+0.1750$i$ & 7.2253 & 2.3593 \\
 0.0819$i$ & 0.5000+0.2646$i$ & 0.3650+0.1750$i$ & 7.6562 & 2.4135 \\
 0.6350+0.1750$i$ & 0.2350+0.1636$i$ & 0.7650+0.1636$i$ & 7.6989 & 2.5260 \\
 0.3650+0.1750$i$ & 0.5000+0.0636$i$ & 0.1176$i$ & 8.1122 & 2.4718 \\
 0.5000+0.1164$i$ & 0.8952+0.1750$i$ & 0.3651+0.1750$i$ & 9.2246 & 2.6469 \\
 0.6351+0.1750$i$ & 0.1052+0.1750$i$ & 0.5000+0.1164$i$ & 9.2304 & 2.6479 \\
 0.6349+0.1750$i$ & 0.5000+0.0572$i$ & 0.1048+0.1750$i$ & 9.5490 & 2.6825 \\
 0.1052+0.1750$i$ & 0.3649+0.1750$i$ & 0.5000+0.0572$i$ & 9.5566 & 2.6838 \\
 0.1050+0.1750$i$ & 0.5000+0.0572$i$ & 0.5000+0.1163$i$ & 11.0593 & 2.8735 \\
 0.5000+0.1130$i$ & 0.5000+0.0551$i$ & 0.3650+0.1750$i$ & 26.4301 & 7.1123 \\
\hline
\end{tabular}
\caption{Numerical solutions of BAE (\ref{BAE;XYZ;M}) and the entanglement NESS spectrum. Here $\eta =0.47$, $N=5$, $M=3$, $\tau =0.35 i$, $\{x_l,y_l\}=\{0.13,0.21\}$.}\label{Tab;XYZ2}
\end{table}

When $\tau\to +i\infty$, the XYZ model reduces to the XXZ model with $\{J_x,J_y,J_z\}\to\{1,1,\cos(\pi\eta)\}$. By letting $iy_{l,r}=\frac{\tau}{2}$ and $\tau\to +i\infty$, the targeted boundary polarizations become
\begin{align}
&n_{l,r}^x = -\frac{ \bell{2}(\frac{\tau}{2})}{\bell{3}(\frac{\tau}{2})} \frac{ \bell{1}(x_{ l,r})}{\bell{4}(x_{l,r})}=-\sin(\pi x_{l,r}), \no\\
&n_{l,r}^y = - i \frac{ \bell{1}(\frac{\tau}{2})}{\bell{3}(\frac{\tau}{2})} \ \frac{ \bell{2}(x_{ l,r})}{\bell{4}(x_{l,r})}=\cos(\pi x_{l,r}), \no\\  
&n_{l,r}^z =- \frac{ \bell{4}(\frac{\tau}{2})}{\bell{3}(\frac{\tau}{2})}\ \frac{ \bell{3}(x_{ l,r})}{\bell{4}(x_{l,r})}=0.
\end{align}
The corresponding BAE (\ref{BAE;XYZ;M}) reduce to the trigonometric ones  \cite{Cao2003,2021PhantomBetheAnsatz,Nepomechie2003}
\begin{align}
&\left[\frac{\sin(\pi u_j+\frac{\pi\eta}{2})}{\sin(\pi u_j-\frac{\pi\eta}{2})}\right]^{2N+2}\left[\frac{\cos(\pi u_j -\frac{\pi\eta}{2})}{\cos(\pi u_j +\frac{\pi\eta}{2})}\right]^2\no\\
&=\prod_{\sigma=\pm1}\prod_{k\neq j}^M\frac{\sin(\pi u_j+\sigma \pi u_k+\pi\eta)}{\sin(\pi u_j+\sigma\pi u_k-\pi\eta)},\qquad j=1,\ldots,M.\label{BAE;XYZ;M2}
\end{align}

In conclusion, we retrieve the result of the previous section in the limit where $iy_{l,r}=\frac{\tau}{2}$ and $\tau\to +i\infty$.

\end{document}